\documentclass[11pt]{article}
\usepackage{graphicx}
\usepackage{subfigure}
\begin{document}
\title{\bf Unitarity of black hole evaporation in final-state
projection models}
\author{Seth Lloyd\\
{\em \small Department of Mechanical Engineering, MIT, Cambridge MA 02139, USA}\\ \\
John Preskill\\
{\em \small Institute for Quantum Information and Matter, Caltech, Pasadena CA 91125, USA}}
\date{}
\maketitle

\begin{abstract}
Almheiri {\em et al.} have emphasized that otherwise reasonable beliefs about black hole evaporation are incompatible with the monogamy of quantum entanglement, a general property of quantum mechanics. We investigate the final-state projection model of black hole evaporation proposed by Horowitz and Maldacena, pointing out that this model admits cloning of quantum states and polygamous entanglement, allowing unitarity of the evaporation process to be reconciled with smoothness of the black hole event horizon. Though the model seems to require carefully tuned dynamics to ensure exact unitarity of the black hole S-matrix, for a generic final-state boundary condition the deviations from unitarity are exponentially small in the black hole entropy; furthermore observers inside black holes need not detect any deviations from standard quantum mechanics.  Though measurements performed inside old black holes could potentially produce causality-violating phenomena, the computational complexity of decoding the Hawking radiation may render the causality violation unobservable. Final-state projection models illustrate how inviolable principles of standard quantum mechanics might be circumvented in a theory of quantum gravity.
\end{abstract}

\section{Introduction}

The quantum physics of black holes has caused great puzzlement since Stephen Hawking discovered \cite{hawking1} nearly 40 years ago that black holes evaporate. The crux of the puzzle is this: if a pure quantum state collapses to form a black hole, the geometry of the evaporating black hole contains spacelike surfaces crossed by both the collapsing body inside the event horizon and nearly all of the emitted Hawking radiation outside the event horizon. If this process is unitary, then the quantum information encoded in the collapsing matter must also be encoded (perhaps in a highly scrambled form) in the outgoing radiation; hence the infalling quantum state is {\em cloned} in the radiation, violating the linearity of quantum mechanics.

This puzzle has spawned many audacious ideas, beginning with Hawking's bold proposal \cite{hawking2} that unitarity fails in quantum gravity. Efforts to rescue unitary led to the formulation of black hole complementarity \cite{complement1,complement2}, the notion that the inside and outside of a black hole are not really two separate subsystems of a composite quantum system, but rather two complementary views of the same system, related by a complex nonlocal map. Black hole complementary set the stage for the holographic principle \cite{holographic1,holographic2}, and its eventual realization in AdS/CFT duality \cite{ads-cft}, which provides a pleasingly unitary picture of black hole evaporation in asymptotically AdS spacetimes, though the implications of this duality regarding the black hole interior remain unclear.

Black hole complementarity seeks to reconcile three reasonable beliefs: (1) An evaporating black hole scrambles quantum information without destroying it. (2) A freely falling observer encounters nothing unusual upon crossing the event horizon of a black hole. (3) An observer who stays outside a black hole detects no violations of relativistic effective quantum field theory. But Almheiri {\em et al.} (AMPS) recently argued \cite{amps} that these three assumptions are incompatible. They consider the Hawking radiation $B$ emitted by a black hole which is nearly maximally entangled with an exterior system $R$. (For example, $R$ could be the radiation so far emitted by an old black hole which has already radiated away more than half of its initial entropy \cite{page}.) Assumptions (1) and (3) require $B$ to be highly entangled with a subsystem $R_B$ of $R$, while assumption (2) requires $B$ to be highly entangled with a subsystem $A$ in the black hole interior. Taken together, then, the three assumptions violate the principle of monogamy of entanglement \cite{terhal,koashi}, which asserts that if quantum systems $A$ and $B$ are maximally entangled, then neither can be correlated with any other system. This tension between unitarity and monogamy had been noted earlier in \cite{mathur,braunstein}.

\begin{figure}[t]
\begin{center}
\includegraphics[width=0.4\textwidth]{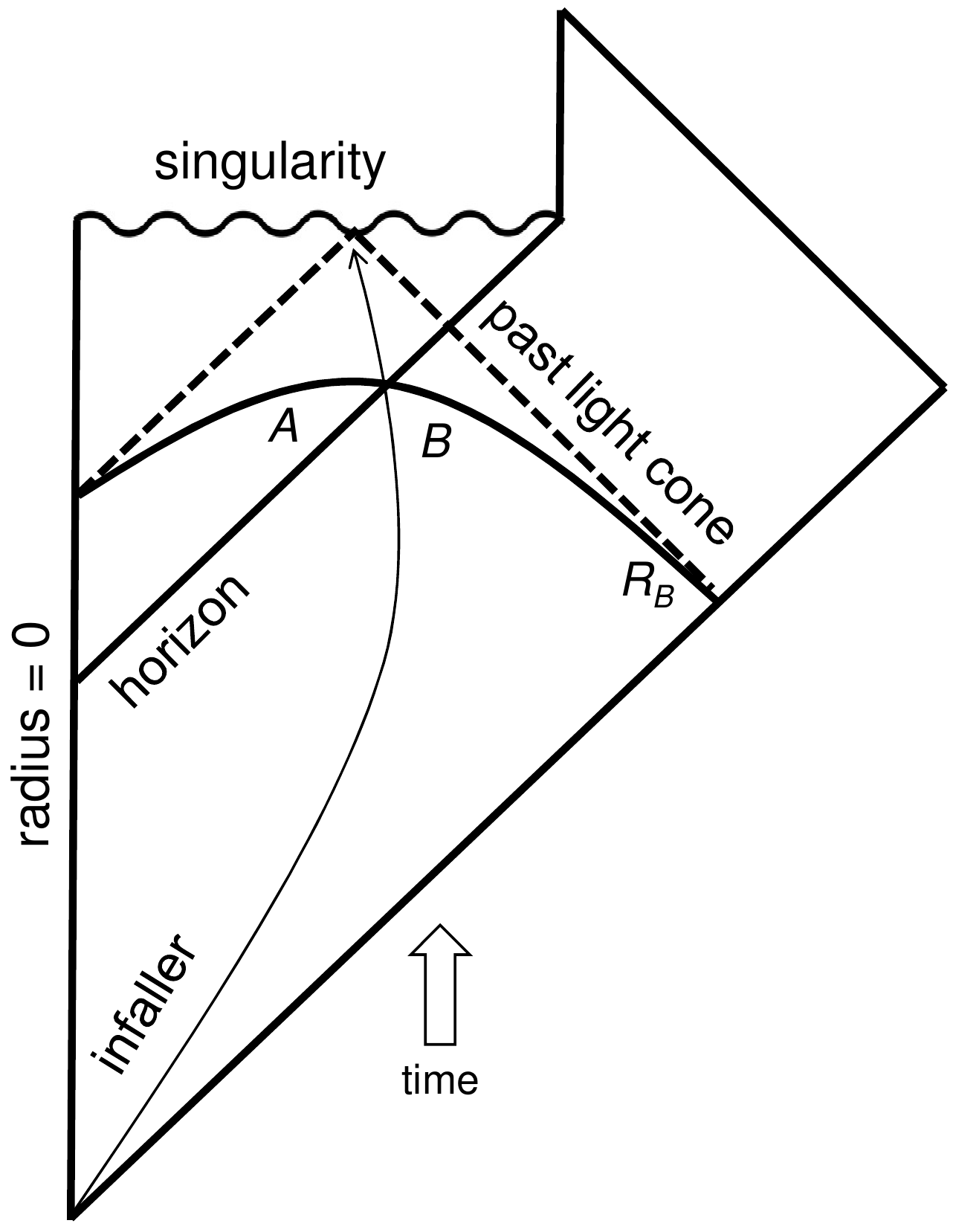}
\end{center}
\caption{The AMPS puzzle: An infalling observer, while still a safe distance from the singularity, is in causal contact with each of the systems $A$, $B$, and $R_B$. If $B$ were highly entangled with both $A$ and $R_B$, the observer should be able to verify the violation of entanglement monogamy.}
\label{fig:amps}
\end{figure}

When the notion of black hole complementarity was initially formulated, it was argued that the cloning of quantum states occurring during black hole evaporation could not be verified in any conceivable experiment, at least within the domain of applicability of trustworthy semiclassical approximations. This observation bolstered the contention that the cloning is operationally fictitious \cite{complement2,preskill,hayden}. In contrast (as indicated in Fig.~\ref{fig:amps}), a single observer falling into the black hole, when still a safe distance from the singularity, could be in causal contact with all three of the systems $A$, $B$, and $R_B$. If monogamy of entanglement were really violated, there is no obvious reason why this observer could not verify the violation. The key difference between the cloning verification experiment described in  \cite{complement2,preskill,hayden} and the entanglement verification described by AMPS \cite{amps}, is that in the former case the observer needs to wait for quantum information to be revealed in the Hawking radiation, by which time it is too late to catch up with its putative clone behind the horizon, while in the latter case no such delay makes system $A$ inaccessible. Thus black hole complementarity needs to be reconsidered.

Like Hawking's original black hole information loss puzzle, the AMPS puzzle has also spawned audacious ideas. AMPS themselves advocated relaxing assumption (2), arguing that an old black hole (and perhaps also a young one) has a singular horizon (a {\em firewall}) and no interior \cite{amps,ampss}. Another possibility is that modifications of assumption (3) allow the entanglement of $B$ with $A$ to be transferred to entanglement of $B$ with $R_B$ as $B$ propagates away from the black hole \cite{giddings}. Or, clinging to a revised version of the complementarity principle, one can assert that $R_B$ should be regarded as a complementary description of $A$ \cite{raju,verlinde}, possibly connected to the black hole interior via a wormhole \cite{maldacena}. All of these ideas will need to be fleshed out further before they can be accurately assessed.

Here we suggest another possible response to the AMPS puzzle, based on the final-state projection model of black hole evaporation proposed \cite{horowitz} by Horowitz and Maldacena (HM). In this scenario, the S-matrix relating the asymptotic incoming state of the collapsing matter and asymptotic outgoing state of the emitted radiation can be unitary; however unitarity can be temporarily violated during the black hole evaporation process, accommodating violations of monogamy of entanglement and the no-cloning principle \cite{wootters,dieks}, and allowing assumptions (1), (2), and (3) to be reconciled. A type of black hole complementarity is realized, and there is no need for firewalls. Just as with other proposed ways to resolve the AMPS puzzle, the HM proposal requires further development before it can be fairly assessed, but we do not regard it as {\em a priori} much more outlandish than these other proposals.

HM proposed imposing a final-state boundary condition requiring a particular quantum state at the spacelike singularity inside the black hole, which allows information to escape from the black hole interior by {\em postselected teleportation}. Speaking fancifully, information residing in the collapsing matter propagates from past infinity to the spacelike singularity inside the black hole, where it is scrambled and reflected, then propagates backward in time from the singularity to the horizon, and forward in time from the horizon to future infinity. More concretely, HM consider the composite system $\mathcal{H}_M\otimes \mathcal{H}_{\rm in}\otimes\mathcal{H}_{\rm out}$, where $\mathcal{H}_M$ is the Hilbert space of the infalling matter, $\mathcal{H}_{\rm in }$ is the Hilbert space of infalling negative energy Hawking radiation behind the horizon, and $\mathcal{H}_{\rm out}$ is the Hilbert space of outgoing positive energy Hawking radiation outside the horizon. What appears to be the vacuum to a freely falling observer crossing the horizon is a maximally entangled state of $\mathcal{H}_{\rm in}\otimes\mathcal{H}_{\rm out}$, and the HM boundary condition projects onto a particular maximally entangled state of $\mathcal{H}_M\otimes\mathcal{H}_{\rm in}$, which encodes the black hole S-matrix. While the horizon crosser sees nothing out of the ordinary, an observer who stays outside the black hole finds that the state of the infalling matter and the state of the outgoing radiation are related by a unitary map.

The HM proposal has the appealing feature that the new physics responsible for evading information loss occurs at the singularity, where we expect semiclassical physics to fail badly. Furthermore, if we are willing to impose initial-state boundary conditions at spacelike singularities in cosmological spacetimes \cite{hartle-hawking}, it may not be unreasonable to impose final-state boundary conditions at spacelike singularities in black hole spacetimes as well. But the proposal has other less pleasing features \cite{gottesman,lloyd}. In particular, unless appropriate constraints are imposed on the dynamics, postselected quantum mechanics can be afflicted with effective closed timelike curves and other causality paradoxes \cite{aharonov,lloyd-ctc1,lloyd-ctc2}; the dynamics may need to be carefully adjusted to protect the unitarity and causality of the evaporation process. 

Our attitude is that these potential bugs in the HM proposal may actually be welcome, helping to steer us toward a deeper understanding of quantum gravity. Therefore, we focus on delineating sufficient conditions for the proposal to work. In brief, we find that the evaporation process is unitary if the interactions between $\mathcal{H}_{\rm in}$ and other systems are appropriately tuned. A deeper understanding of quantum gravity may be needed to decide whether black hole evaporation really fulfills these conditions, but we find that for a generic final-state boundary condition at the singularity, the deviations from exact unitarity scale like $e^{-S_{BH}/2}$ where $S_{BH}$ is the black hole entropy. Such exponentially small violations of unitarity could well be regarded as a success for our semiclassical analysis of the HM model, since nonperturbative quantum gravity corrections of that order are expected and are beyond the scope of the analysis. We also argue, again assuming a generic final-state boundary condition, that deviations from standard quantum theory are unlikely to be detected by infalling observers as they approach the singularity. 

A careful analysis of the entanglement-verifying measurements discussed by AMPS reveals that imposing unitarity of the black hole S-matrix may not suffice to ensure that the HM model is physically sensible; in particular measurements performed inside an old black hole might produce causality violating modifications of the Hawking radiation outside the black hole, enabling observers who stay outside to send signals backward in time.  This backward signaling can be achieved, however, only by rapidly decoding the Hawking radiation; hence the computational complexity of this task \cite{harlow-hayden} may suffice to enforce causality.

Even if it turns out that the HM model is not realized in nature, the model is still quite instructive. It cautions us that inviolable consequences of standard quantum mechanics, such as the no-cloning principle and monogamy of entanglement, need not be respected in quantum gravity. Perhaps that is the proper lesson to be drawn from the AMPS puzzle.

\begin{figure}[t]
\begin{center}
\includegraphics[width=0.8\textwidth]{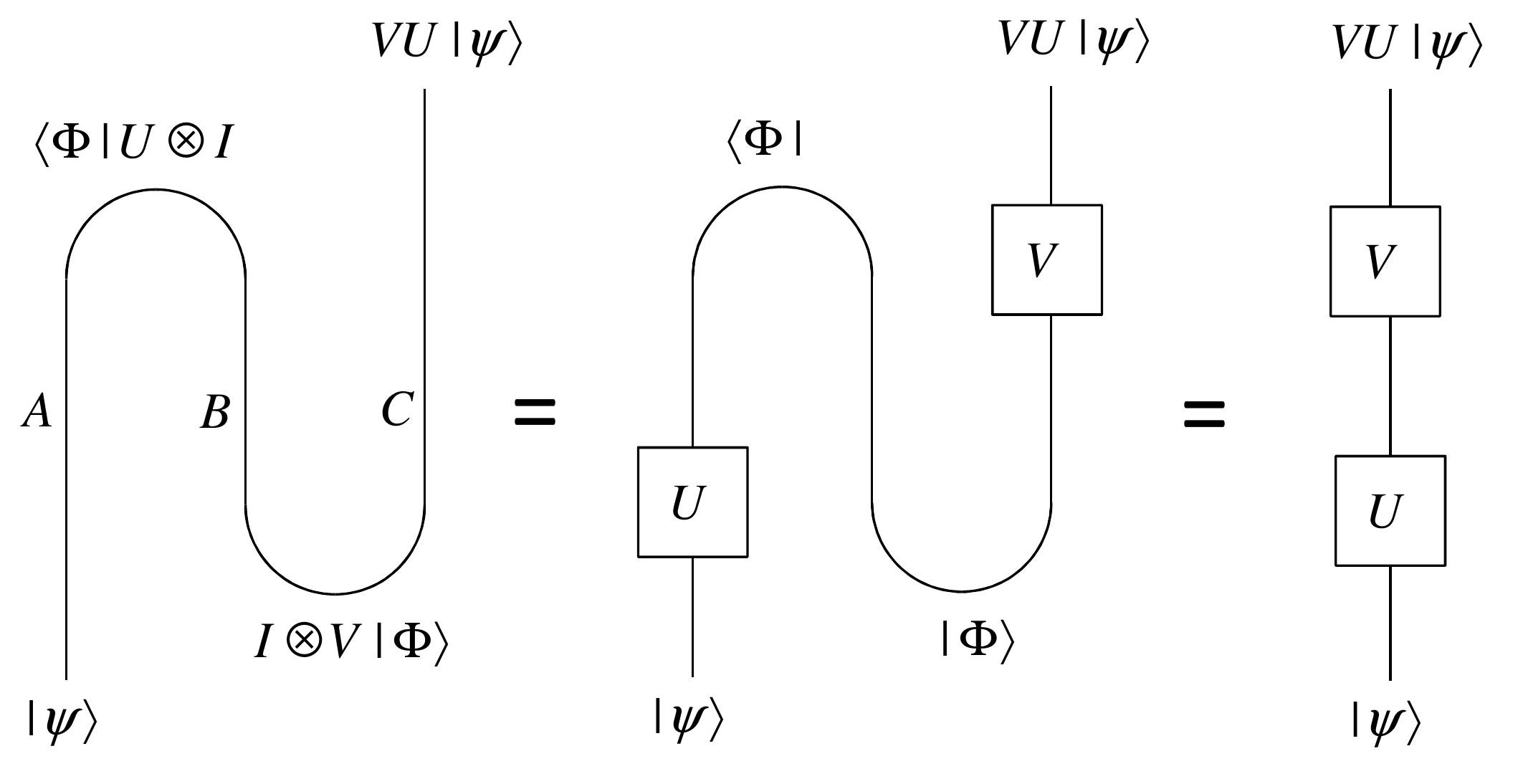}
\end{center}
\caption{Quantum teleportation. To convey a quantum state $|\psi\rangle$ of system $A$ to system $C$, first a maximally entangled state $|\Phi(V)\rangle$ of $BC$ is prepared, and then $AB$ is projected to a maximally entangled state $|\Phi(U^*)\rangle$. To recover $|\psi\rangle$, a party at $C$ applies the unitary transformation $U^\dagger V^\dagger$.}
\label{fig:teleport}
\end{figure}

After reviewing the HM model in Sec.~2, we comment on its relevance to the AMPS controversy in Sec.~3. In Sec.~4 we examine how unitarity of the black hole S-matrix might fail in the HM model, concluding that, for a generic final-state boundary condition at the singularity, the deviations from exact unitarity are exponentially small in the black hole entropy. In Sec.~5 we argue that observers with limited access to the infalling Hawking radiation need not detect any deviations from standard quantum mechanics. We examine measurements performed inside a black hole in Sec.~6, discussing in particular whether such measurements can enable acausal signaling. Sec.~7 contains some concluding comments.

Connections between the AMPS puzzle and the HM model have also been discussed in \cite{verlinde,bousso,bousso-stanford,rangamani}.

\section{The Horowitz-Maldacena proposal}

\begin{figure}[t]
\begin{center}
\includegraphics[width=0.9\textwidth]{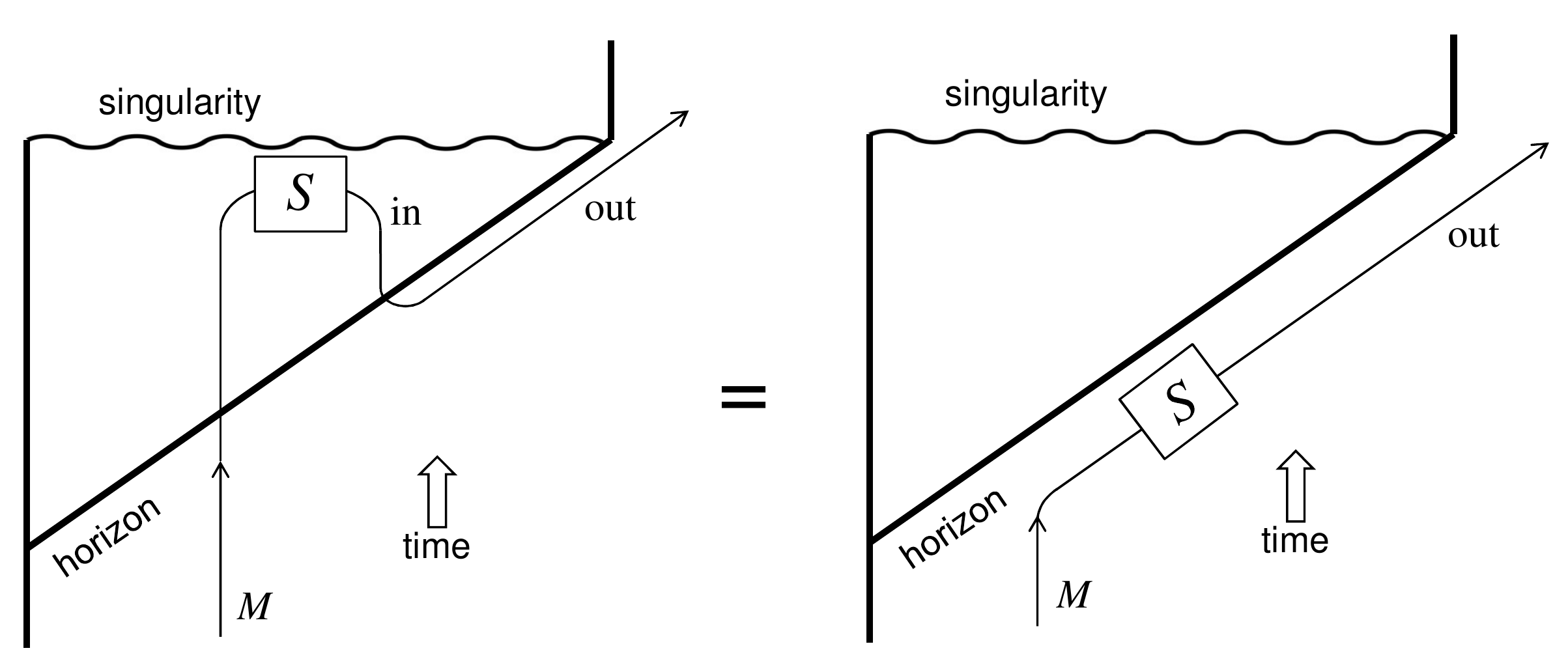}
\end{center}
\caption{The Horowitz-Maldacena model, in which quantum information carried by the collapsing matter system $M$ is teleported out of a black hole. Outgoing Hawking radiation is maximally entangled with infalling radiation, and a final-state boundary condition projects $M$ and the infalling radiation to a maximally entangled state which encodes the unitary S-matrix $S$.  }
\label{fig:penrose}
\end{figure}

The HM proposal is based on quantum teleportation \cite{teleportation}, which is illustrated in Fig.~\ref{fig:teleport}. Any maximally entangled pure state of two $d$-dimensional systems $A$ and $B$ can be expressed as 
\begin{eqnarray}
|\Phi(V)\rangle\equiv \left(I\otimes V \right)|\Phi\rangle= \left(V^T\otimes I\right)|\Phi\rangle.
\end{eqnarray}
Here $|\Phi\rangle = \frac{1}{\sqrt{d}}\sum_{i=1}^d |i\rangle_A\otimes |i\rangle_B$, where $\{|i\rangle_A\}$, $\{|i\rangle_B\}$ denote orthonormal bases, $V$ is a unitary $d\times d$ matrix, and $V^T$ is the transpose of $V$. To teleport the state $|\psi\rangle$ from $A$ to $C$, we first prepare the entangled state $|\Phi(V)\rangle_{BC}$ of system $BC$, then perform an entangled measurement on $AB$. If the outcome of the measurement is $|\Phi(U^*)\rangle$, then up to normalization the state of $C$ becomes
\begin{eqnarray}
\left({}_{AB}\langle \Phi(U^*) |\Phi(V)\rangle_{BC}\right)|\psi\rangle_A = V_C\left({}_{AB}\langle \Phi |\Phi\rangle_{BC}\right)U_A|\psi\rangle_A = \frac{1}{d}VU|\psi\rangle_C,
\end{eqnarray}
where the factor $1/d$ indicates that the measurement outcome $|\Phi(U^*)\rangle$ occurs with probability $1/d^2$. Once known, this outcome can be transmitted to $C$ by classical communication, and if the initial entangled state of $BC$ is also known, then a party at $C$ can apply $U^\dagger V^\dagger$ to recover the state $|\psi\rangle$ in system $C$. If either the initial state of $BC$ or the projected state of $AB$ were not maximally entangled, then either $V$ or $U$ would be non-unitary and hence unphysical; in that case the teleportation process would have imperfect fidelity.

In the HM proposal depicted in Fig.~\ref{fig:penrose}, quantum information is teleported from the collapsing matter system $\mathcal{H}_M$, the source for the black hole's classical geometry, to the outgoing Hawking radiation system $\mathcal{H}_{\rm out}$ that is emitted as the black hole evaporates. The dimension $d$ is the number of distinguishable microstates for a black hole with specified total mass. Because the final-state boundary condition specifies that only one particular maximally entangled state is accepted at the singularity, there is no need for classical communication to convey the outcome of the entangled measurement.

The initial maximally entangled state used in the protocol is the Unruh state $|\Phi\rangle_{{\rm in}\otimes {\rm out}}$, which looks like the vacuum state to a freely falling observer who crosses the horizon. Here $\mathcal{H}_{\rm in}$ is a system of infalling Hawking quanta behind the horizon. We use a microcanonical description, summing over all microstates with approximately the same energy, so that this state is maximally entangled rather than thermal. (The microcanonical ensemble is appropriate if we wish to consider the formation and evaporation of a black hole with sharply defined energy; of course, an observer with access to a small subsystem of $\mathcal{H}_{\rm out}$ will see a thermal state.) By a suitable basis choice, we set the unitary matrix specifying this maximally entangled state to the identity. 

Loosely speaking, the basis state $|i\rangle_{\rm in}$ of $\mathcal{H}_{\rm in}$ is the negative energy Hawking state behind the horizon paired with the positive energy Hawking state $|i\rangle_{\rm out}$ outside the black hole. ``Negative energy'' is really a misnomer, because the timelike Killing vector of the exterior geometry becomes spacelike behind the horizon; hence ``energy'' inside the black hole is really momentum. In any case this description of the Unruh state is not precise because the evaporating black hole is not static and has no Killing vector. We take it for granted, though, that the notion of a maximally entangled state of $\mathcal{H}_{\rm in}\otimes \mathcal{H}_{\rm out}$ can be made precise.

If the entangled state of $\mathcal{H}_M\otimes \mathcal{H}_{\rm in}$ specified by the final-state boundary condition is  $|\Phi(S^*)\rangle_{M\otimes {\rm in}}$, where $S$ is unitary, then the infalling matter state and the outgoing radiation state are related by $|\varphi\rangle_{\rm out} = S |\psi\rangle_M$; thus $S$ is the black hole S-matrix, presumed to be a highly nonlocal scrambling unitary transformation. $S$ is required to rigorously satisfy conservation of energy and other exact gauge charges, but it need not respect global symmetries, which are expected to be broken in quantum gravity. 

Though analytically extended non-Schwarzschild black hole geometries can have timelike rather than spacelike singularities, the interior geometries of these solutions are unstable \cite{poisson}, and we assume the singularity is always spacelike and unavoidable in realistic collapse scenarios. Because the final-state boundary condition accepts any quantum state of the infalling matter system, observers approaching the singularity, particularly those with access to only a local subsystem, need not experience a reversal in the arrow of time or any departure from the usual laws of quantum mechanics. We will discuss this point further in Sec.~\ref{sec:detecting}.

\section{Features of the model}
\label{sec:features}

As Fig.~\ref{fig:penrose} indicates, the HM model supports a characteristic flow of information in spacetime, which ensures the unitarity of the black hole evaporation process. Information initially encoded in the collapsing matter flows forward in time from past infinity to the spacelike singularity, then backward in time from the singularity to the horizon, and finally forward in time from the horizon to future infinity. Despite the apparently acausal propagation backward in time, there is an equivalent description of the same process with a conventional causal ordering; the information flow can be ``pulled tight'' to ``straighten out'' the bends in the flow. This alternative description can be strictly justified only if the infalling radiation system $\mathcal{H}_{\rm in}$ is perfectly isolated from $\mathcal{H}_{\rm out}$ and $\mathcal{H}_M$, which may not be precisely true; therefore In Sec.~\ref{sec:conditions} we will revisit the sufficient conditions for unitarity in a more general setting. But for now we will assume that the information flow admits a consistent causal ordering, and consider some of the consequences.

\begin{figure}
\begin{center}
\includegraphics[width=0.7\textwidth]{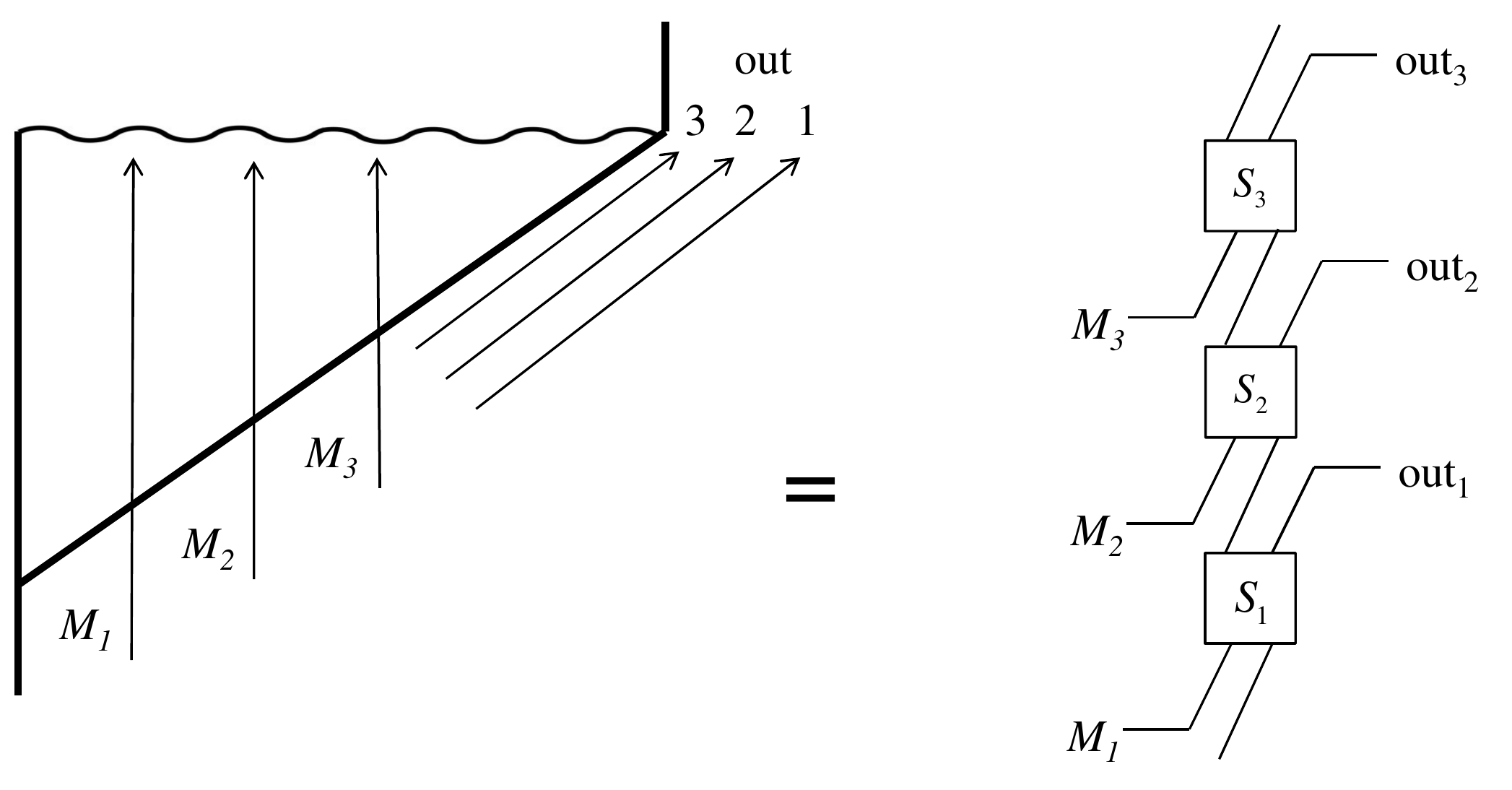}
\end{center}
\caption{Information flow for a black hole that maintains its mass by accreting a steady stream of infalling matter.}
\label{fig:feed}
\end{figure}

\subsection{Relaxing the no-cloning principle: black hole complementarity}
Once straightened, the overall process clearly preserves quantum information, with the unitary matrix $S$ appearing in the final-state boundary condition playing the roll of the S-matrix relating the asymptotic incoming and outgoing states. But at intermediate times anomalous phenomena can occur, which would be disallowed in standard unitary quantum mechanics. For example, as Fig.~\ref{fig:penrose} illustrates, cloning of quantum states can occur in postselected quantum mechanics. The quantum information encoded in $\mathcal{H}_M$ is also available, albeit in a highly scrambled form, in $\mathcal{H}_{\rm out}$ on the same spacelike slice. From the perspective of the causally ordered straightened process, the cloned state in the outgoing radiation is merely the same as the state of the infalling matter, except viewed at a later ``time'' and in a different basis.

Thus black hole complementarity is realized in the HM model in the sense that observables inside and outside the horizon acting on the same spacelike slice do not commute. From the perspective of the causally ordered information flow, this failure of commutativity is expected, because the outside observables act on the same system as the inside observables, but at a later ``time.''

We may also consider a process in which we continually feed a black hole with additional matter to maintain its mass for a long time compared to its natural evaporation time, before finally allowing the evaporation to proceed to completion. In that case the S-matrix $S$, rather than being an arbitrary unitary transformation mapping the infalling matter to the outgoing radiation, must have a special structure enforced by the requirement that the entropy of the radiation should never exceed the Bekenstein-Hawking entropy of the black hole, if the overall state is pure. The information processing can be described by a quantum circuit whose bounded width is determined by the black hole entropy as in Fig.~\ref{fig:feed}, and in particular the final-state boundary condition will respect the requirement that information cannot escape from the evaporating black hole before it falls in. If this circuit scrambles rapidly \cite{sekino}, then the ``information mirror'' phenomenon \cite{hayden} will occur, in which, for a black hole highly entangled with its surroundings, information absorbed by the black hole returns in the emitted radiation after a Schwarzschild time $O( m \log m)$, where $m$ is the black hole mass. We note that if additional mass is thrown into the black hole after it initially forms, to avoid firewalls we require a smooth Unruh vacuum at the apparent horizon, not at the global horizon whose position depends on the future history of the hole. 

\subsection{Relaxing entanglement monogamy: easing the AMPS puzzle}

\begin{figure}[t]
\begin{center}
\includegraphics[width=0.8\textwidth]{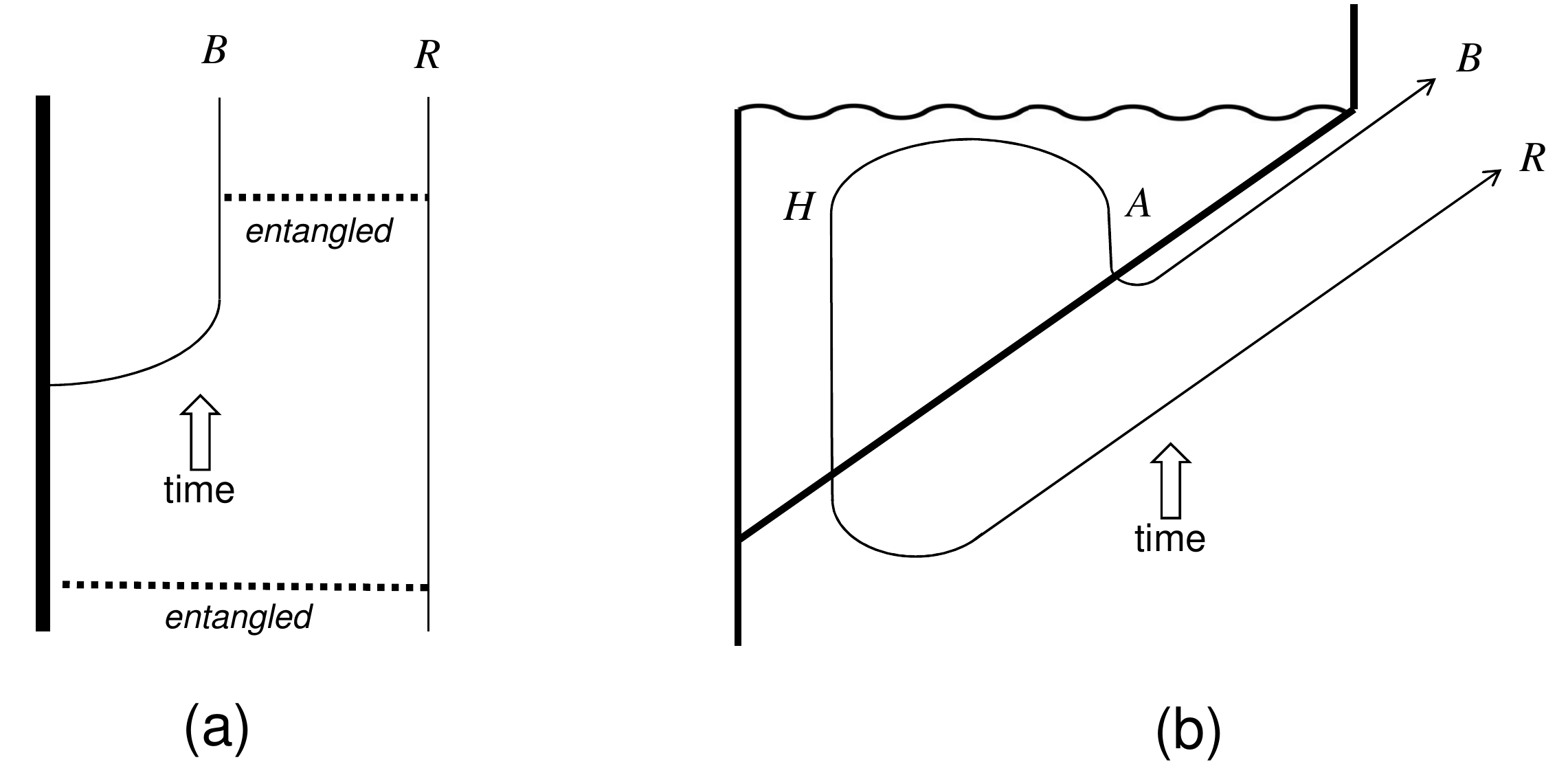}
\end{center}
\caption{(a) An old black hole is maximally entangled with system $R$ in the previously emitted Hawking radiation. After the black hole emits the Hawking quanta $B$, $R$ is entangled with $B$ and no longer entangled with the black hole. (b) Entanglement transfer in the HM model. $AB$ and $HR$ are maximally entangled Unruh vacuum states. The final-state projection of $HA$ onto a maximally entangled state creates maximal entanglement of $BR$ via entanglement swapping.}
\label{fig:transfer}
\end{figure}

Now, following AMPS \cite{amps,ampss}, we consider the case of an ``old''  black hole which has already emitted more than half of its initial entropy, so that its microscopic degrees of freedom have become maximally entangled with its previously emitted radiation \cite{page}. As depicted in Fig.~\ref{fig:transfer}, let $B$ denote some Hawking quanta which have been recently emitted by this old black hole. If the horizon looks smooth to infalling observers, then $B$ should be maximally entangled with system $A$ behind the horizon, the Unruh partners of the $B$ quanta. On the other hand, unitarity of the evaporation process requires $B$ to be maximally entangled with a subsystem $R$ of the previously emitted radiation. In standard quantum mechanics entanglement is monogamous --- $B$ cannot be entangled with each of the two systems $A$ and $R$. To relieve this tension AMPS proposed that $B$ is entangled with $R$ but not $A$; hence the infalling observer encounters a firewall at the horizon. 

In the HM model, however, smoothness of the horizon can be reconciled with unitarity, as shown in Fig.~\ref{fig:transfer}. We note by $H$ the Unruh partner of $R$ inside the black hole; hence both $AB$ and of $HR$ are maximally entangled Unruh vacuum states. Now, for a particular fixed state of the infalling matter system $M$, the final-state boundary condition at the singularity projects the infalling radiation onto the corresponding state. If we suppose that the boundary condition projects $HA$ onto a maximally entangled state, then the resulting postselected state of $BR$ is maximally entangled as well. This phenomenon is called {\em entanglement swapping} \cite{swapping1,swapping2}. In standard entanglement swapping, an entangled measurement is performed on $HA$, and the outcome of this measurement must be communicated to $BR$ to complete the swapping protocol. No such communication is necessary in the HM model, because the boundary condition dictates that only one possible outcome can occur. In principle an observer outside the black hole could successfully verify the $BR$ entanglement, while an infalling observer could pass safely through the horizon, verifying the $AB$ entanglement.

This picture is oversimplified. For one thing, we have ignored the computational complexity of extracting $R$ from the Hawking radiation \cite{harlow-hayden}, which must be achieved in order to verify the $BR$ entanglement. Furthermore, measurements in postselected quantum mechanics raise some daunting conceptual puzzles. We will return to these issues in Sec.~\ref{sec:measurements}.

\section{Conditions for unitarity}
\label{sec:conditions}

The discussion in Sec.~\ref{sec:features} was premised on the assumption that the postselected information flow in spacetime has a consistent causal ordering, ensuring the unitarity of the evaporation process. In general, though, interactions among the systems $\mathcal{H}_M$, $\mathcal{H}_{\rm in}$ and $\mathcal{H}_{\rm out}$ might disrupt this ordering; will the black hole S-matrix still be unitary in that case?

\subsection{Entanglement across the horizon}

One important criterion for unitarity concerns the entangled state of $\mathcal{H}_{\rm in}\otimes \mathcal{H}_{\rm out}$ which is used as a resource in postselected teleportation.

Assuming $|\mathcal{H}_{\rm in}| = |\mathcal{H}_{\rm out}| = d$ (where $|\mathcal{H}|$ denotes the dimension of $\mathcal{H}$) we say that an entangled state of $\mathcal{H}_{\rm in}\otimes \mathcal{H}_{\rm out}$ is ``full rank'' if the marginal density operator on $\mathcal{H}_{\rm in}$ (and hence also  $\mathcal{H}_{\rm out}$) has $d$ nonzero (possibly degenerate) eigenvalues. Any full-rank bipartite entangled state can be expressed as $\left(U\otimes I \right)|\Phi\rangle$, where $|\Phi\rangle$ is a canonical maximally entangled state, and $U$ is invertible (though not necessarily unitary). 

\begin{figure}[t]
\begin{center}
\includegraphics[width=0.5\textwidth]{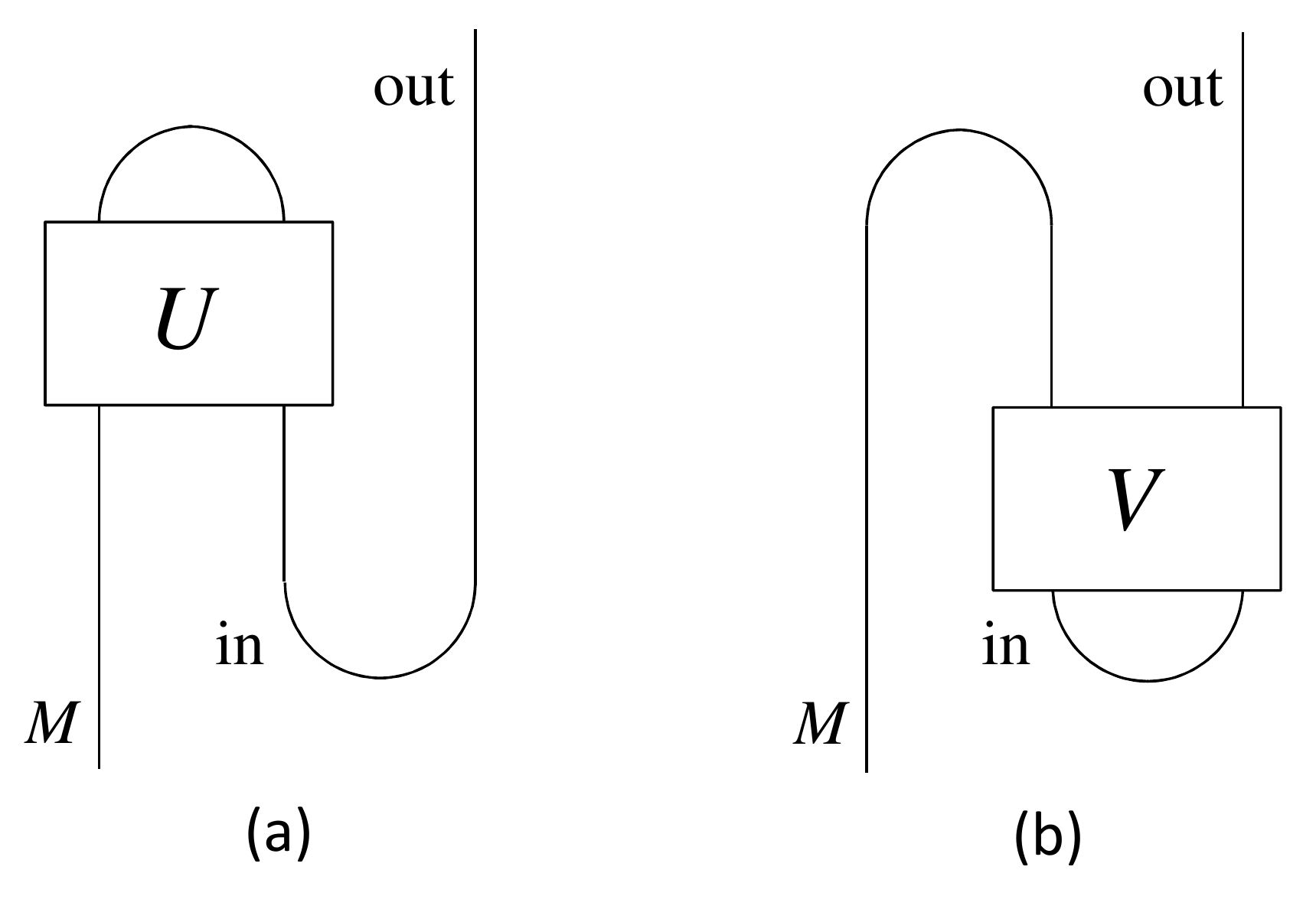}
\end{center}
\caption{Entangling interactions between the infalling radiation and the collapsing matter (a), or between the infalling radiation and the outgoing radiation (b), may compromise the fidelity of postselected teleportation.}
\label{fig:entangling}
\end{figure}

If the initial state used in postselected teleportation is  the full-rank entangled state $|\Psi\rangle_{{\rm in}\otimes {\rm out}}=\left(I\otimes U\right)|\Phi\rangle_{{\rm in}\otimes {\rm out}}$, and the final-state boundary condition projects onto $ {}_{M\otimes{\rm in}}\langle\Theta |={}_{M\otimes{\rm in}}\langle \Phi|\left(U^{-1}S\otimes I\right)$, where $S$ is unitary, then the black hole S-matrix will be $S$. Neither the initial state nor the postselected state is maximally entangled, but the non-maximal entanglement of the $\langle \Theta|$ compensates perfectly for the non-maximally entanglement of $|\Psi\rangle$, resulting in overall unitarity. We see that the state at the apparent horizon need not be maximally entangled to ensure the unitarity of postselected teleportation, as long as the final-state condition is adjusted appropriately.

However, as we will discuss in Sec.~\ref{subsec:generic} below, it seems natural to conjecture that the postselected state is in some sense generic, which means that ${}_{M\otimes{\rm in}}\langle \Theta|$ is likely to be very close to maximally entangled. In that case, unitarity demands that $|\Psi\rangle_{{\rm in}\otimes {\rm out}}$ be very nearly maximally entangled as well.

We have another reason to demand a high degree of entanglement for the state $|\Psi\rangle_{{\rm in}\otimes {\rm out}}$: a freely falling observer crossing the apparent horizon should see a smooth vacuum state rather than a seething firewall. Smoothness at the horizon requires the state to closely resemble the Unruh state, in which a mode localized outside the horizon which has sharply defined frequency with respect to Schwarzschild time is entangled with its Unruh partner behind the horizon, such that the reduced density operator of either mode is thermal when its partner is traced out. If the state that collapses to form a black hole has nearly definite energy, then we presume that the reduced density operator on  $\mathcal{H}_{\rm out}$ for the global state of this Unruh vacuum is nearly maximally mixed --- it is essentially the microcanonical ensemble in a narrow energy band, whose purification is a nearly maximally entangled state on $\mathcal{H}_{\rm in}\otimes \mathcal{H}_{\rm out}$. Actually, the compatibility of the mode-by-mode thermal entanglement (required for smoothness of the horizon) with the near maximal entanglement of the global state (required for unitarity) is a delicate quantitative issue which we find hard to resolve decisively; related issues were discussed in \cite{marolf}. For most of the rest of our discussion, we will just assume that the initial state of $\mathcal{H}_{\rm in}\otimes \mathcal{H}_{\rm out}$ is maximally entangled, though in Sec.~\ref{sec:measurements} we will revisit how this entanglement is affected by horizon-crossing agents who interact with both $\mathcal{H}_{\rm out}$ and $\mathcal{H}_{\rm in}$.

\subsection{A generic final state}
\label{subsec:generic}
Unitarity may fail due to interactions between $\mathcal{H}_{\rm in}$ and the other systems. In the teleportation circuit, quantum information effectively flows backward in time in $\mathcal{H}_{\rm in}$, and interactions of such chronology violating systems with chronology respecting systems can be dangerous, inducing closed timelike curves, and hence failure of unitarity \cite{lloyd-ctc1,lloyd-ctc2}. Put more prosaically, entangling interactions behind the horizon between $\mathcal{H}_{\rm in}$ and $\mathcal{H}_M$, as in Fig.~\ref{fig:entangling}a, compromise the fidelity of teleportation, because in effect $\mathcal{H}_M\otimes\mathcal{H}_{\rm in}$ will not be projected onto a maximally entangled state. Likewise, entangling interactions between $\mathcal{H}_{\rm in}$ and $\mathcal{H}_{\rm out}$, as in Fig.~\ref{fig:entangling}b, also cause trouble because in effect the state of $\mathcal{H}_{\rm in}\otimes\mathcal{H}_{\rm out}$ used in the teleportation protocol will not be maximally entangled. 

Let's assume that the state of $\mathcal{H}_{\rm in}\otimes \mathcal{H}_{\rm out}$ is exactly maximally entangled, and consider the consequences of entangling interactions between $\mathcal{H}_{\rm in}$ and $\mathcal{H}_M$ behind the horizon, as in Fig.~\ref{fig:entangling}a. Intriguingly, if the final-state projection is chosen generically, or equivalently if the unitary transformation $U$ in Fig.~\ref{fig:entangling}a acting on $\mathcal{H}_M\otimes\mathcal{H}_{\rm in}$ is sampled uniformly with respect to the invariant Haar measure, then the evaporation process is very, very nearly, though not quite exactly, unitary.

A black hole with mass $m$ has entropy $O(m^2)$ and evaporation time $O(m^3)$. The vast majority of ways of making a black hole look like the time-reversed evaporation process and require a time $O(m^3)$. Black holes created rapidly, in time $O(m)$, have entropy $O(m^{3/2})$, and hence have many fewer possible microstates than generic black holes. Analysis of the creation and evaporation of a generic black hole may be subtle, because substantial evaporation occurs while the black hole is still being assembled. Let's focus instead on the case where the black hole forms rapidly. We divide the Hilbert space of the infalling matter into two subsystems, $\mathcal{H}_M=\mathcal{H}_{M_1}\otimes \mathcal{H}_{M_2}$, where the states in $\mathcal{H}_{M_1}$ collapse rapidly; hence $|\mathcal{H}_{M_1}|/|\mathcal{H}_{M}|= \exp\left(-O(m^2)\right)\ll 1$, where $|
\mathcal{H}|$ denotes the dimension of the Hilbert space $\mathcal{H}$.

\begin{figure}[t]
\begin{center}
\includegraphics[width=0.3\textwidth]{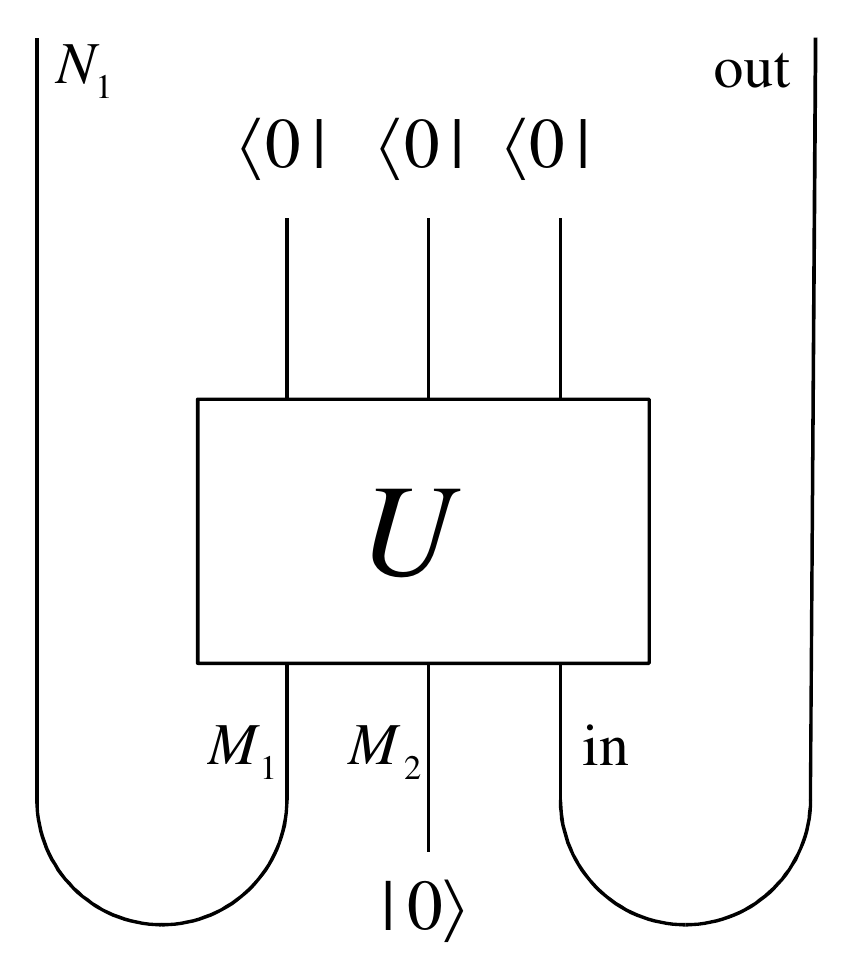}
\end{center}
\caption{The HM model for a generic final-state boundary condition. Subsystem $M_1$ of the collapsing matter system $M$ is maximally entangled with a reference system $N_1$, and $\mathcal{H}_{M}\otimes \mathcal{H}_{\rm in}$ is projected onto a Haar-random state determined by the unitary transformation $U$. In the resulting postselected state, $N_1$ is very nearly maximally entangled with a subsystem of the outgoing Hawking radiation.}
\label{fig:generic}
\end{figure}

For the purpose of analyzing whether quantum information initially carried by the rapidly collapsing matter system $\mathcal{H}_{M_1}$ can be decoded from the outgoing Hawking radiation $\mathcal{H}_{\rm out}$, it is convenient to ask what happens when $M_1$ is maximally entangled with a reference system $N_1$ as shown in Fig.~\ref{fig:generic}. We assume that subsystem $M_2$ starts out in a fixed state, {\em e.g.}, its vacuum state. After the final-state projection, a random pure state $|\Psi(U)\rangle_{N_1\otimes{\rm out}}$ on $\mathcal{H}_{N_1}\otimes \mathcal{H}_{\rm out}$ is obtained, which depends on the unitary transformation $U$ that defines the postselected state of $\mathcal{H}_{M}\otimes \mathcal{H}_{\rm in}$. Tracing out the radiation system we obtain a mixed marginal state $\rho_{N_1}(U)$ on $N_1$, and by averaging over $U$ we find \cite{decoupling}
\begin{eqnarray}
\int dU \|\rho_{N_1}(U)-\rho_{N_1}^{\rm max}  \|_1 \le \sqrt{\frac{|\mathcal{H}_{M_1}|}{|\mathcal{H}_{\rm in}|}}\approx \exp\left(-S_{BH}/2+O(m^{3/2})\right);
\end{eqnarray}
here $\|\cdot\|_1$ denotes the $L^1$-norm, $dU$ is the normalized Haar measure on the unitary group, $\rho_{N_1}^{\rm max}$ is the maximally mixed state on $N_1$, and $S_{BH}=\ln |\mathcal{H}_{\rm in}|$ is the black hole entropy. Thus the typical state on $N_1$ is extremely close to maximally mixed.

Since the overall state of $\mathcal{H}_{N_1}\otimes\mathcal{H}_{\rm out}$ is pure, that $\rho_{N_1}$ is almost maximally mixed means that the reference system $N_1$ is almost maximally entangled with a subsystem of the outgoing Hawking radiation, and correspondingly that a unitary decoding map acting on $\mathcal{H}_{\rm out}$ can isolate this subsystem which almost purifies $\rho_{N_1}$. It follows that for a Haar-typical final-state projection, an arbitrary initial state of $M_1$ can be decoded in the outgoing Hawking radiation with a fidelity deviating from one by just $\exp\left(-O(m^2)\right)$. A similar conclusion would still apply if the unitary $U$ were sampled from a unitary 2-design rather than the Haar measure, a sampling task which (unlike sampling from Haar measure) can be achieved exactly by a relatively small quantum circuit with size $O(m^4)$, or approximately with error $\epsilon$ by circuits with depth $O\left(\log m\log(1/\epsilon)\right)$ \cite{cleve}. 

Nearly perfect unitarity is gratifying, but exact unitarity is what we yearn for. To ensure exact unitarity, we must restrict the form of the initial and final entangled states in the HM model, as well as the interactions of $\mathcal{H}_{\rm in}$ with infalling matter behind the horizon. This necessary fine-tuning in the model has been criticized \cite{gottesman,lloyd}, but one might instead regard it as a tantalizing hint about quantum gravitational dynamics. Surely, that generic final-state projections come so close to achieving unitarity enhances the plausibility of the dynamical constraints we demand. Violations of unitarity scaling like $e^{-S_{BH}/2}$ could well be artifacts of the semiclassical framework used in the formulation of the HM model, as nonperturbative quantum gravitational corrections of that order are expected. Furthermore, information loss at such a tiny scale would be exceedingly difficult to detect ``in practice,'' even if we disregard the complexity of decoding the highly scrambled Hawking radiation \cite{harlow-hayden}. Indeed, the deviation from exact unitarity might be undetectable even in principle until the very last stage of the black hole evaporation process, when semiclassical methods no longer apply. Since assuming a generic final-state boundary condition is just a rather crude guess, finding such an excellent approximation to exact unitarity might be regarded as a success rather than a failure of the HM model.

\section{Detecting postselection when approaching the singularity}
\label{sec:detecting}

Up until now we have focused on the unitarity of the black hole S-matrix relating the asymptotic infalling matter and the asymptotic outgoing radiation. Even if this S-matrix is exactly unitary, though, infalling observers inside the black hole might still experience departures from conventional quantum theory in the HM model, arising from entangling interactions between $\mathcal{H}_M$ and  $\mathcal{H}_{\rm in}$. What do infalling observers see?

Entangling interactions of $\mathcal{H}_M$ and  $\mathcal{H}_{\rm in}$ could be induced by the Hamiltonian dynamics as the matter falls from the horizon to the singularity. If so, these interactions must be suitably augmented or reversed by the final-state projection in order to ensure unitarity of the black hole S-matrix. The resulting information flow behind the horizon does not have a well defined causal order, or in other words if we try to define a causal order we find that the quantum information encoded in the time-reversed infalling radiation could in principle interact with its earlier self encoded in the collapsing matter. 

\begin{figure}[t]
\begin{center}
\includegraphics[width=0.3\textwidth]{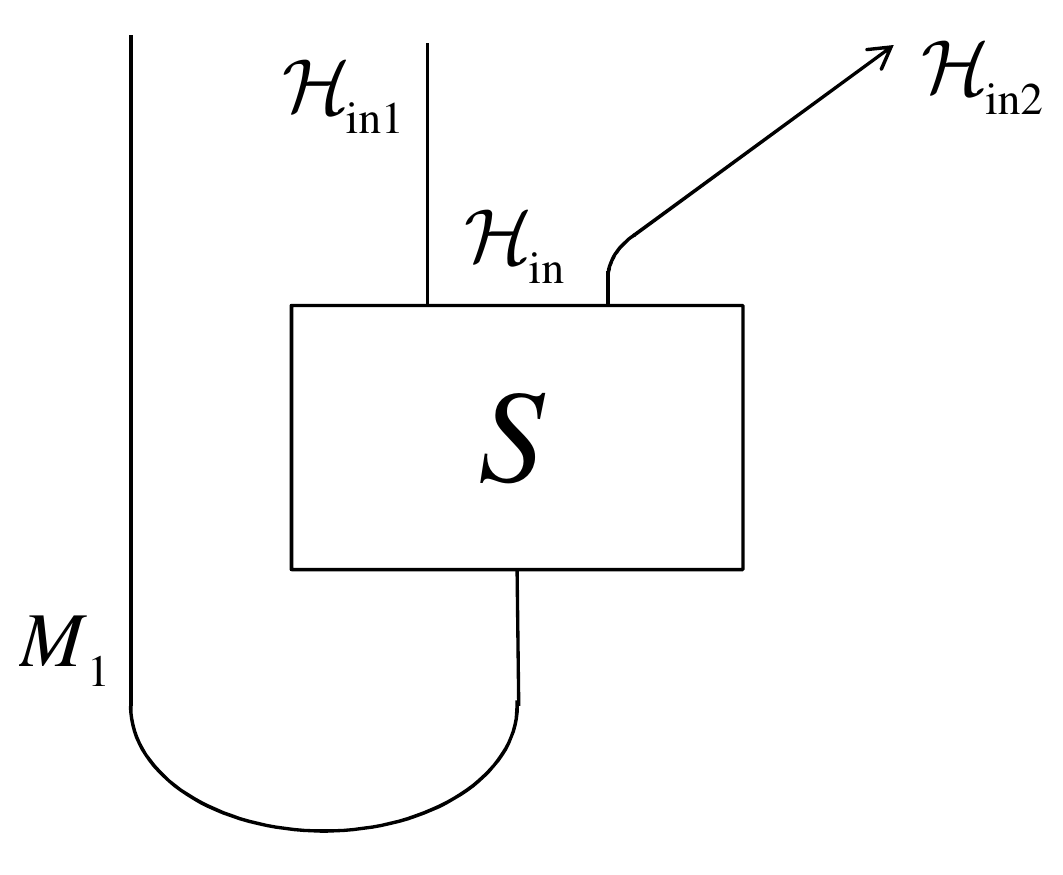}
\end{center}
\caption{The collapsing matter subsystem $M_1$ is maximally entangled with a subsystem of the infalling radiation system $\mathcal{H}_{\rm in}$, specified by the scrambling unitary transformation $S$. If the subsystem $\mathcal{H}_{{\rm in}2}$ of $\mathcal{H}_{\rm in}$ is discarded, then the complementary subsystem $\mathcal{H}_{{\rm in}1}$ becomes nearly uncorrelated with $M_1$, if $\mathcal{H}_{{\rm in}2}$ is larger than half the size of $\mathcal{H}_{M_1}\otimes \mathcal{H}_{\rm in}$.}
\label{fig:discard}
\end{figure}

On the other hand, because the final-state projection throughly scrambles the quantum information encoded in the collapsing matter, the weird consequences of such closed timelike curves behind the horizon may be undetectable by infalling observers with access to only a portion of the $\mathcal{H}_{\rm in}$ Hilbert space. To clarify this claim, consider Fig.~\ref{fig:discard}, which depicts the time-reversed evolution from the singularity into the black hole interior. Here $M_1$ is a subsystem of the collapsing matter, which is maximally entangled with a subsystem, determined by the scrambling unitary $S$, of the infalling radiation system $\mathcal{H}_{\rm in}$. Suppose we discard the subsystem $\mathcal{H}_{{\rm in}2}$ of $\mathcal{H}_{\rm in}$, presumed inaccessible to our infalling observer, and retain the complementary subsystem $\mathcal{H}_{{\rm in}1}$, which the observer might be able to access. Averaging $S$ over the normalized invariant Haar measure on the unitary group, and assuming that the overall state of $\mathcal{H}_{M_1}\otimes \mathcal{H}_{\rm in}$ is pure, we find that the density operator $\rho_{M_1,{\rm in1}}$ obeys the inequality \cite{decoupling}
\begin{eqnarray}
\int dS \|\rho_{M_1,{\rm in1}}(S)-\rho_{M_1}(S)\otimes \rho_{\rm in1}^{\rm max}  \|_1 \le \sqrt{\frac{|\mathcal{H}_{M_1}|\cdot |\mathcal{H}_{\rm in1}|}{|\mathcal{H}_{\rm in2}|}},
\end{eqnarray}
where $\rho_{\rm in1}^{\rm max}$ denotes the maximally entangled state of $\mathcal{H}_{{\rm in}1}$. The conclusion is that, for generic $S$, if the discarded system $\mathcal{H}_{{\rm in}2}$ is larger than half the full system $\mathcal{H}_{M_1}\otimes \mathcal{H}_{\rm in}$, then $M_1$ is hardly entangled with $\mathcal{H}_{{\rm in}1}$ at all; instead it is nearly maximally entangled with $\mathcal{H}_{{\rm in}2}$. Specifically, if $\log_2|M_1| = k$, $\log_2|\mathcal{H}_{\rm in}| = n$, and $\log_2 |\mathcal{H}_{{\rm in}1}|= \frac{1}{2}(n-k)-r$, we find that the state of $M_{1,{\rm in 1}}$ deviates in the $L^1$-norm from an uncorrelated product state by at most $2^{-r}$. As in Sec.~\ref{subsec:generic}, we obtain the same result by averaging over a unitary 2-design rather than Haar measure.

Translated into the language of the HM model, this statement means that when quantum information encoded in a small subsystem of the collapsing matter Hilbert space is ``reflected'' at the singularity by a generic final-state boundary condition, the reflected information escapes the notice of an observer with access to much less than half of the infalling radiation. An infalling observer who crosses the event horizon of a black hole with mass $m$ meets the singularity in proper time $O(m)$, and hence has very limited time to perform complex decoding operations on the infalling radiation. This observer may suffer horribly when subjected to the highly nonlocal scrambling transformation $S$ at the singularity, but she might not have time to discern any other troubling violations of the rules of standard quantum mechanics.


\section{Measurements inside black holes}
\label{sec:measurements}

So far we have argued that the HM model yields unitary (or very nearly unitary) black hole dynamics under generic conditions. But the AMPS puzzle concerns infalling observers who perform highly nongeneric measurements. To address more fully how the AMPS puzzle is resolved by the HM model, we must examine more deeply how the final-state boundary condition affects measurements performed inside a black hole, or measurements which straddle the black hole horizon. 

In standard measurement theory, we usually suppose that the measured system interacts unitarily with a suitable ``meter'' and that subsequent interactions of the meter with its environment cause the measurement alternatives to decohere in a particular basis. The meter provides a record which can be consulted later on to verify the outcome of the measurement. 

In the HM model we can imagine a meter (which can be regarded as a late-arriving component of the infalling matter system $\mathcal{H}_M$), which falls into the black hole and interacts with the infalling radiation system $\mathcal{H}_{\rm in}$. The meter eventually reaches the singularity, and is subjected to the final-state boundary condition. Does this mean that the record of the measurement outcome is destroyed?

Not necessarily. If the overall evolution is unitary, then the meter, like the rest of the infalling matter, will be teleported out of the black hole. A relic of the meter survives in the outgoing Hawking radiation, albeit in a highly scrambled form. In principle, the scrambled meter can be extracted and decoded by performing a complex quantum computation. Indeed, this decoding of the meter might be done long after the black hole has evaporated completely and disappeared. In this sense, a record of the measurement survives, at least in principle, which can be consulted later on to verify the measurement outcome, just as in standard measurement theory. 

We have seen that unitary interactions inside the black hole between $\mathcal{H}_M$ and $\mathcal{H}_{\rm in}$ can threaten the unitarity of the black hole S-matrix. One way to protect unitarity is to demand that such interactions are in effect undone by the final-state boundary condition at the singularity. In that case, however, records of measurements performed inside the black hole would be permanently erased, and could not be extracted from the Hawking radiation after the black hole has evaporated. We might then question the operational meaning of such measurements. And if the degrees of freedom inside the black hole cannot be measured, even in principle, in what sense can the black hole interior be said to exist?

As we will discuss, though, measurements performed inside black holes which are not permanently erased can be reconciled with unitarity of the black hole S-matrix if the interactions between system and meter obey suitable constraints. In that case, however, an agent behind the horizon may be able to send causality-violating signals into her backward light cone; furthermore, this causality violation may be detectable by observers who remain outside the black hole. Such causality violation in the bulk spacetime might be hard to reconcile with a dual boundary description of the dynamics which is unitary and causal; a possible resolution is that the computational complexity of decoding the Hawking radiation \cite{harlow-hayden} prevents backward signaling, hence protecting causality.

It seems that either point of view --- (1) that measurements performed inside black hole are undone by the final-state condition, or (2) that measurement outcomes can be recovered by decoding the Hawking radiation --- raises intriguing questions about the viability of the HM model. 

\begin{figure}[t]
\begin{center}
\includegraphics[width=0.9\textwidth]{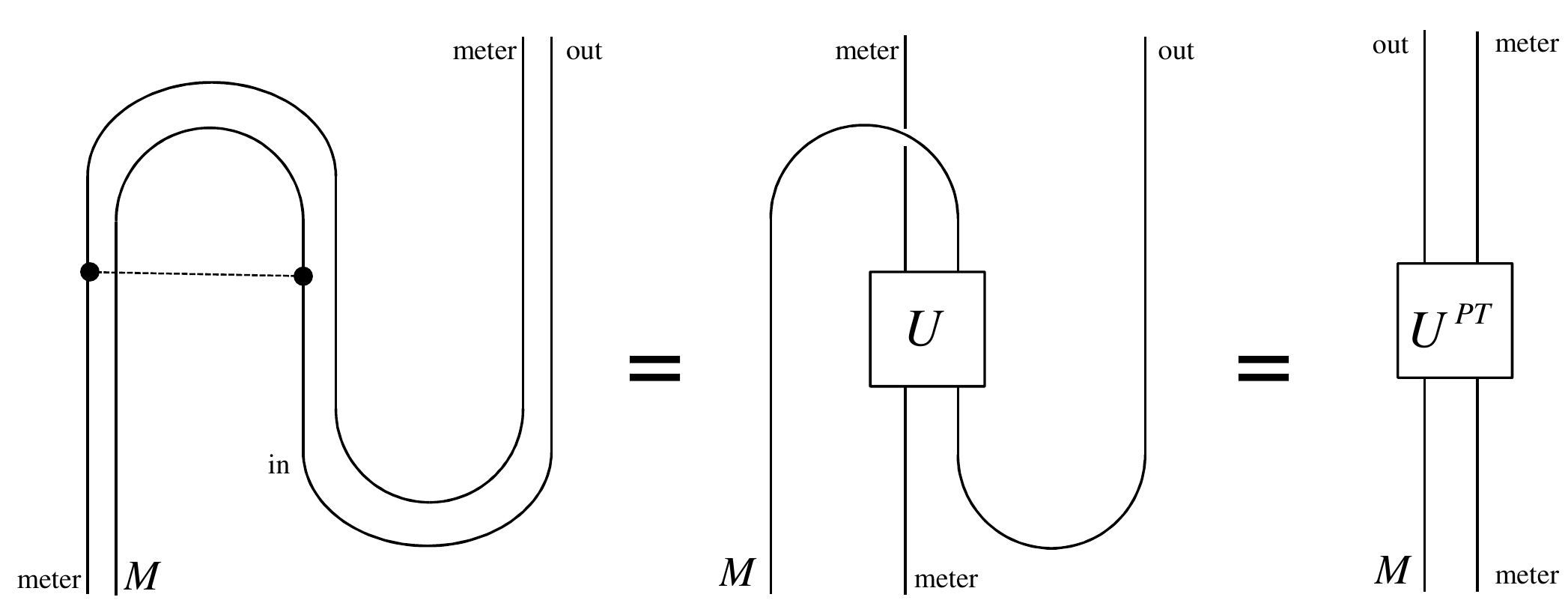}
\end{center}
\caption{Measurement of the infalling Hawking radiation inside a black hole. The meter used in the measurement interacts with the measured system behind the horizon; then the meter is teleported out of the black hole. In the equivalent causally order process, the unitary interaction $U$ between system and meter is replaced by its partial transpose $U^{\rm PT}$.}
\label{fig:inside-measure}
\end{figure}

\subsection{Unitarity constraint}

Consider a measurement of the infalling Hawking radiation behind the horizon. A meter is prepared outside the horizon and dropped into the black hole. It is programmed to interact with a subsystem of $\mathcal{H}_{\rm in}$; an entangling unitary transformation $U$ is applied to the meter and that subsystem as in Fig.~\ref{fig:inside-measure}. At the singularity, the meter is paired with a different subsystem of $\mathcal{H}_{\rm in}$ and teleported out of the black hole.

As indicated in Fig.~\ref{fig:inside-measure}, this process is equivalent to one in which the operator $U^{\rm PT}$ acts on the meter and the infalling matter subsystem that is paired with the measured subsystem of $\mathcal{H}_{\rm in}$. Here ``PT'' denotes ``partial transpose,'' meaning that the initial and final states of the $M$-out system are transposed, while those of the meter are not. In general, the partial transpose of a unitary operator is not unitary, and therefore the resulting S-matrix is not unitary. But if $U^{\rm PT}$ is unitary, then the transformation $U$ acting on $\mathcal{H}_{\rm in}$ behind the horizon is compatible with unitarity of the S-matrix. This is just a special case of the criterion for unitarity of the S-matrix formulated in \cite{gottesman}.

For example, suppose that $U$ is the entangling unitary that realizes an orthogonal measurement performed on the infalling radiation:
\begin{equation}
U = \sum_{a,b} \Pi^a_{\rm in} \otimes \left(|b+a\rangle\langle b|\right)_{\rm meter}.
\end{equation}
Here $\{\Pi_{\rm in}^a, a=1,2,3,\dots N\}$ is a complete set of orthogonal projectors acting on the measured subsystem, and $\{|b\rangle_{\rm meter}, b = 1,2,3,\dots N\}$ is an orthonormal basis for the meter. Hence the state of the meter shifts by $a$ if the measured state is in the support of $\Pi_a$ (the addition $b+a$ is modulo $N$). The partial transpose of this unitary,
\begin{equation}
U = \sum_{a,b} \left(\Pi^a_{M}\right)^T \otimes \left(|b+a\rangle\langle b|\right)_{\rm meter},
\end{equation}
is also unitary, since $\{\left(\Pi_{M}^a\right)^T, a=1,2,3,\dots N\}$ is also a complete set of orthogonal projectors. We see, therefore, that orthogonal measurements performed on $\mathcal{H}_{\rm in}$ inside the black hole need not violate unitarity. If such measurements are to be forbidden, it must be on some other grounds. 

\subsection{Chronology violation}

\begin{figure}[t]
\begin{center}
\includegraphics[width=0.3\textwidth]{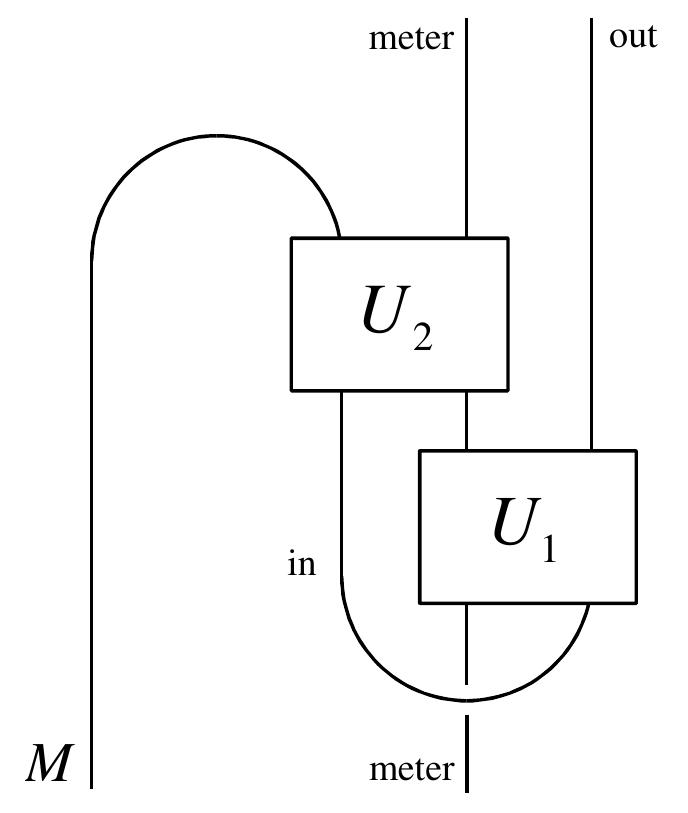}
\end{center}
\caption{A measurement straddling the black hole horizon. From the meter's perspective the interaction $U_1$ occurs first and is followed by $U_2$, while from the perspective of the radiation the interactions occur in the opposite order. }
\label{fig:straddle-measure}
\end{figure}

We also wish to consider measurements which straddle the black hole horizon. In such a measurement, a meter interacts first with a subsystem of $\mathcal{H}_{\rm out}$ (the unitary transformation $U_1$), then falls into the black hole and interacts with $\mathcal{H}_{\rm in}$ (the unitary transformation $U_2$), as shown in Fig.~\ref{fig:straddle-measure}.

In this case, even if $U_2$ has a unitary partial transpose, the resulting circuit has a peculiar property --- it is {\em chronology violating}. There is no equivalent circuit, without postselection, which describes the information flow and admits a globally defined forward time direction for both the radiation and the meter. The information flow cannot be ``pulled tight'' because from the meter's perspective the interaction $U_1$ occurs first and is followed by $U_2$, while from the perspective of the radiation the interactions occur in the opposite order. 

\begin{figure}[t]
\begin{center}
\includegraphics[width=0.9\textwidth]{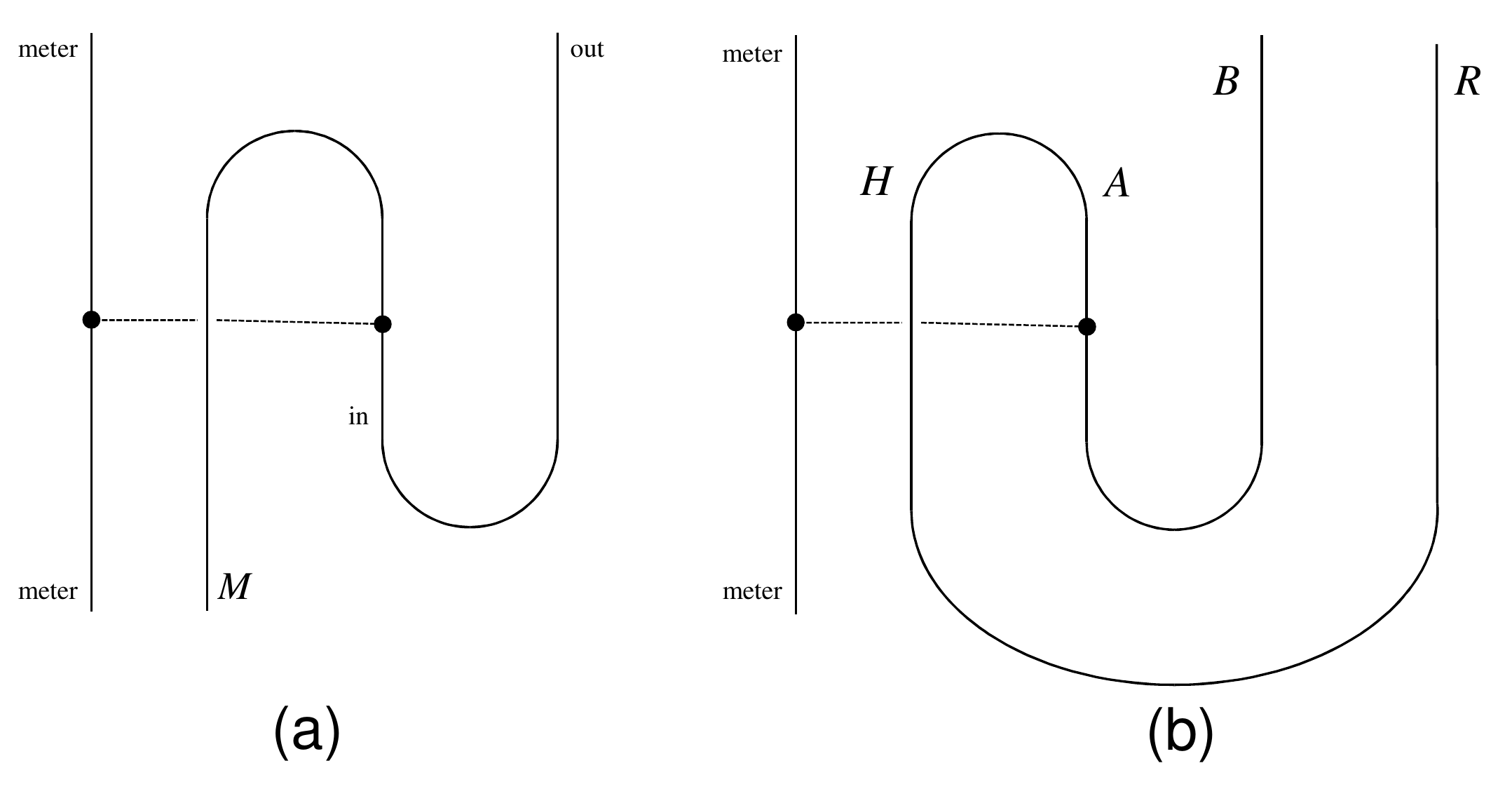}
\end{center}
\caption{(a) A measurement inside a young black hole may influence the Hawking radiation outside, but the effect of the measurement is not immediately detectable. (b) A measurement inside an old black hole modifies the state of $BR$ outside the horizon.}
\label{fig:acausal}
\end{figure}

Even if the S-matrix relating the asymptotic infalling matter to the asymptotic outgoing radiation is unitary, this chronology violation may lead to acausal effects in the bulk spacetime. For example an agent falling into the black hole, by interacting with $\mathcal{H}_{\rm in}$, can influence the state of the Hawking radiation in $\mathcal{H}_{\rm out}$ that was emitted before the agent fell in, as in Fig.~\ref{fig:acausal}a. 

In the case of a young black hole (one not entangled with its surroundings), this ability to influence the Hawking radiation by tossing a meter into the black hole is not so disturbing. The previously emitted Hawking radiation is highly mixed because it is entangled with the black hole microstates. An observer outside the black hole can interpret the process depicted in Fig.~\ref{fig:acausal}a by saying that the meter which falls in becomes rapidly distributed among the black hole's degrees of freedom. Much later, a scrambled version of the meter, having been emitted in the Hawking radiation, could wind up being entangled with the radiation that had already been emitted before the meter fell in.

However for an old black hole (one highly entangled with its previously emitted radiation) as in Fig.~\ref{fig:acausal}b, interactions of an infalling meter with $\mathcal{H}_{\rm in}$ behind the horizon may produce genuinely detectable acausal effects outside the black hole. Here the final-state condition dictates that the recently emitted radiation system $B$ is maximally entangled with the previously emitted system $R$. But if a meter tossed into the black hole interacts with system $A$ behind the horizon, the meter becomes entangled with $BR$, so the state of $BR$ is no longer pure. Thus the outcomes of measurements performed on $BR$ today can depend on whether or not we decide to toss the meter into the black hole tomorrow. We will analyze such causality violating effects in more detail in the following subsections.

\subsection{Entanglement-verifying measurements}

To prepare for the ensuing discussion of the AMPS experiment, we will first discuss in detail how one can use a quantum computer to perform measurements which project onto a maximally entangled basis. For simplicity we will consider measurement of qubit pairs; the discussion can easily be extended to higher dimensional systems. 

Our goal is to measure the qubit pair $AB$ in the Bell basis:
\begin{eqnarray}
|\phi^\pm\rangle_{AB} &=&\frac{1}{\sqrt{2}}\left(|00\rangle_{AB} \pm |11\rangle_{AB}\right),\nonumber\\
|\psi^\pm\rangle_{AB} &=&\frac{1}{\sqrt{2}}\left(|01\rangle_{AB} \pm |10\rangle_{AB}\right).
\end{eqnarray}
These four orthonormal states can be usefully characterized as the simultaneous eigenstates of the two commuting Pauli operators $X_A\otimes X_B$ and $Z_A\otimes Z_B$, where $X$ and $Z$ are the Pauli matrices
\begin{equation}
X = \left( \begin{array}{cc}
0 & 1  \\
1 & 0 \end{array} \right) ,\quad 
Z = \left( \begin{array}{cc}
1 & 0  \\
0 & -1 \end{array} \right).
\end{equation}
$Z\otimes Z$ is the ``parity bit'' of the qubit pair, distinguishing the strings 00 and 11 from the strings 01 and 10. $X\otimes X$ is the ``phase bit,'' distinguishing the $\pm$ superpositions of two strings of the same parity. If a two-qubit state is expected to be $|\phi^+\rangle$, we can verify the state by checking both $X\otimes X = 1$ and $Z\otimes Z=1$. 

\begin{figure}[t]
\begin{center}
\includegraphics[width=0.7\textwidth]{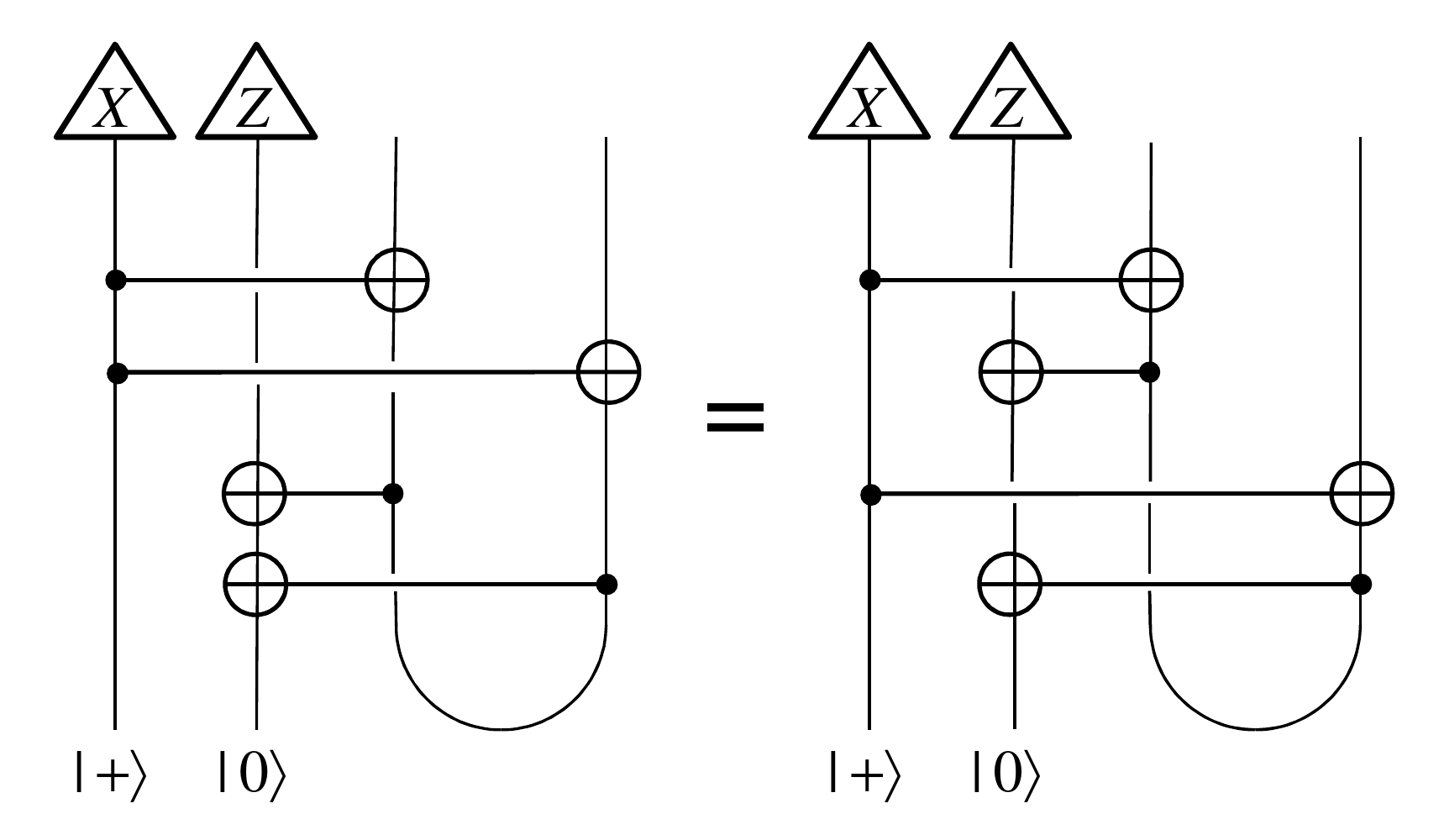}
\end{center}
\caption{Quantum circuit for measuring a qubit pair in the maximally entangled Bell basis. In the circuit on the left, the measurement of $Z\otimes Z$ is performed first, followed by the measurement of $X\otimes X$. In the equivalent circuit on the right, the meter interacts first with one of the two measured qubits, then with the other. This latter circuit may be used by an infalling agent to verify entanglement that straddles a black hole event horizon.}
\label{fig:bell-meas}
\end{figure}

A quantum circuit for measuring in the Bell basis is shown in Fig.~\ref{fig:bell-meas}. The circuit uses a coherent two-qubit quantum gate, the controlled-NOT (CNOT) gate, whose action on a complete basis is
\begin{eqnarray}
{\rm CNOT}: |a, b\rangle \mapsto |a, a\oplus b\rangle,
\end{eqnarray}
where $a,b, \in \{0,1\}$. Two meter qubits are used in the measurement, one each for the measurement of $Z\otimes Z$ and $X\otimes X$. To measure $Z\otimes Z$, we prepare the meter qubit in the $Z=1$ eigenstate $|0\rangle$, perform two successive CNOT gates from the qubits to be measured to the meter, and then read out the meter by measuring $Z$. Hence we read out the sum modulo two of the two measured qubits without collecting any additional information. To measure $X\otimes X$, we prepare the meter qubit in the $X=1$ eigenstate $|+\rangle = \frac{1}{\sqrt{2}}\left(|0\rangle + |1\rangle\right)$, perform two successive CNOT gates from the meter to the qubits to be measured, and then read out the meter by measuring $X$. The pair of CNOT gates applies $X\otimes X$ to the measured qubits if the meter is in the state $|1\rangle$ (and does nothing if the meter is in the state $|0\rangle$), hence flipping the meter from the $X=1$ eigenstate $|+\rangle$ to the $X=-1$ eigenstate $|-\rangle$ (or not) depending on whether the eigenvalue of $X\otimes X$ is $-1$ or +1.

If qubit $B$ is outside the horizon of a black hole, and qubit $A$ is inside the horizon, then the Bell measurement should be performed sequentially; the meter interacts with $B$ first, then falls through the horizon to interact with $A$. After commuting two CNOT gates as in Fig.~\ref{fig:bell-meas}, our Bell measurement circuit has the desired sequential form. The two-qubit meter falls to the singularity and is teleported out of the black hole. The meter is thoroughly scrambled with other qubits in the Hawking radiation emitted by the black hole, but in principle it can be decoded and consulted later on to verify the entanglement of the qubit pair $AB$.

\subsection{The AMPS experiment}

Now we will discuss entanglement-verifying measurements performed on an old black hole, as considered by AMPS. For a particular pure quantum state of the infalling matter from which the black hole initially formed, $R$ is a subsystem of the early radiation, maximally entangled with the recently emitted radiation $B$. $A$ is the Unruh partner of $B$ behind the horizon, also maximally entangled with $B$, while $H$ is the Unruh partner of $R$. The final-state boundary condition at the singularity inside the black hole projects $HA$ onto a maximally entangled state. For conceptual clarity and notational simplicity, we suppose that $H$, $A$, $B$, and $R$ are all single-qubit systems, that the Unruh state is $|\phi^+\rangle_{HR}\otimes|\phi^+\rangle_{AB}$, and that the boundary condition projects $HA$ onto $|\phi^+\rangle_{HA}$.

\begin{figure}[t]
\begin{center}
\includegraphics[width=0.9\textwidth]{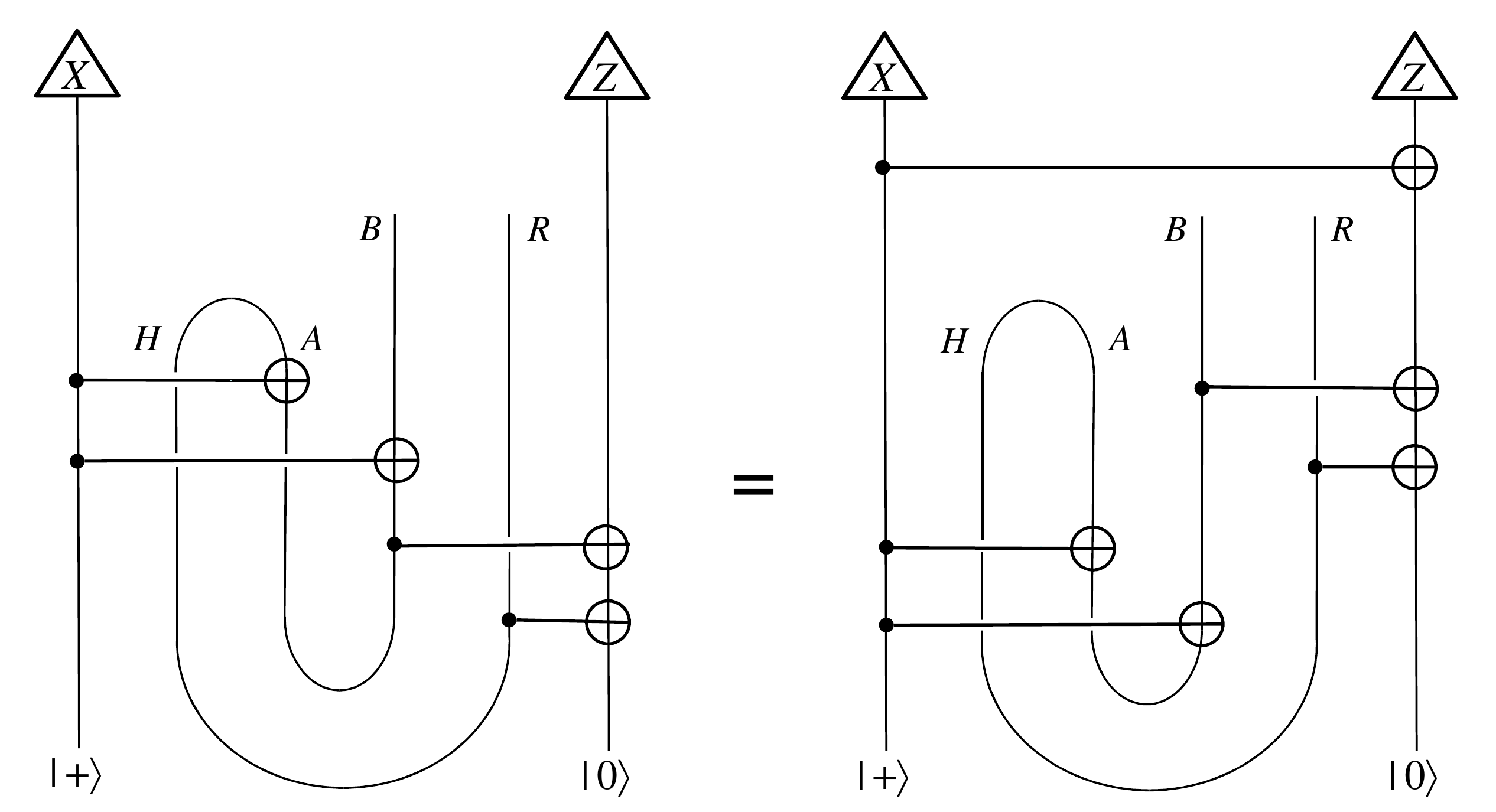}
\end{center}
\caption{Entanglement verifying measurements for an old black hole. In the circuit on the left, BRenda measures $Z_B\otimes Z_R$ before ABby measures $X_A\otimes X_B$. (ABby performs her measurement by interacting with $B$ outside the horizon first, then falling into the black hole to interact with $A$ inside the horizon later.) In the equivalent circuit on the right, ABby measures before BRenda, and there is an additional CNOT gate from ABby's meter to BRenda's.}
\label{fig:brenda-first}
\end{figure}

Let us suppose for now that the system $R$ can be extracted efficiently from the Hawking radiation, despite the complexity of this task  \cite{harlow-hayden}. In principle, then, a single agent could perform an entangled measurement of $BR$ followed by an entangled measurement of $AB$. To illustrate our essential point, though, it will suffice to consider two different agents acting independently: BRenda, who measures $BR$ and ABby, who measures $AB$. BRenda never needs to enter the black hole; ABby interacts with $B$ first, and then falls into the black hole to interact with $A$ later on.

If there were no ABby, BRenda could perform Bell measurement on $BR$, obtaining the outcome $|\phi^+\rangle$ with probability 1, hence successfully verifying the expected $BR$ entanglement. Just to be sure, BRenda could measure $BR$ many times (each time with a fresh meter), obtaining the outcome $|\phi^+\rangle$ every time.

If there we no BRenda, ABby could perform Bell measurement on $AB$, obtaining the outcome $|\phi^+\rangle$ with probability 1, hence successfully verifying the expected $AB$ entanglement. Just to be sure, ABby could measure $AB$ many times (each time with a fresh meter), obtaining the outcome $|\phi^+\rangle$ every time. Each of ABby's meters could be extracted from the Hawking radiation after the black hole has evaporated, allowing us to confirm her results. 

If both BRenda and ABby measure, then both must interact with $B$. Provided ABby interacts with $B$ before BRenda, both BRenda and ABby successfully verify the entanglement as described above. But if BRenda interacts with $B$ before ABby, then neither verification succeeds. To understand what happens, it suffices to suppose that each party performs just half of the Bell measurement --- BRenda measures $Z_B\otimes Z_R$ and ABby measures $X_A\otimes X_B$. These two measurements do not commute, so their order makes a difference. 

\begin{figure}[t]
\begin{center}
\includegraphics[width=0.9\textwidth]{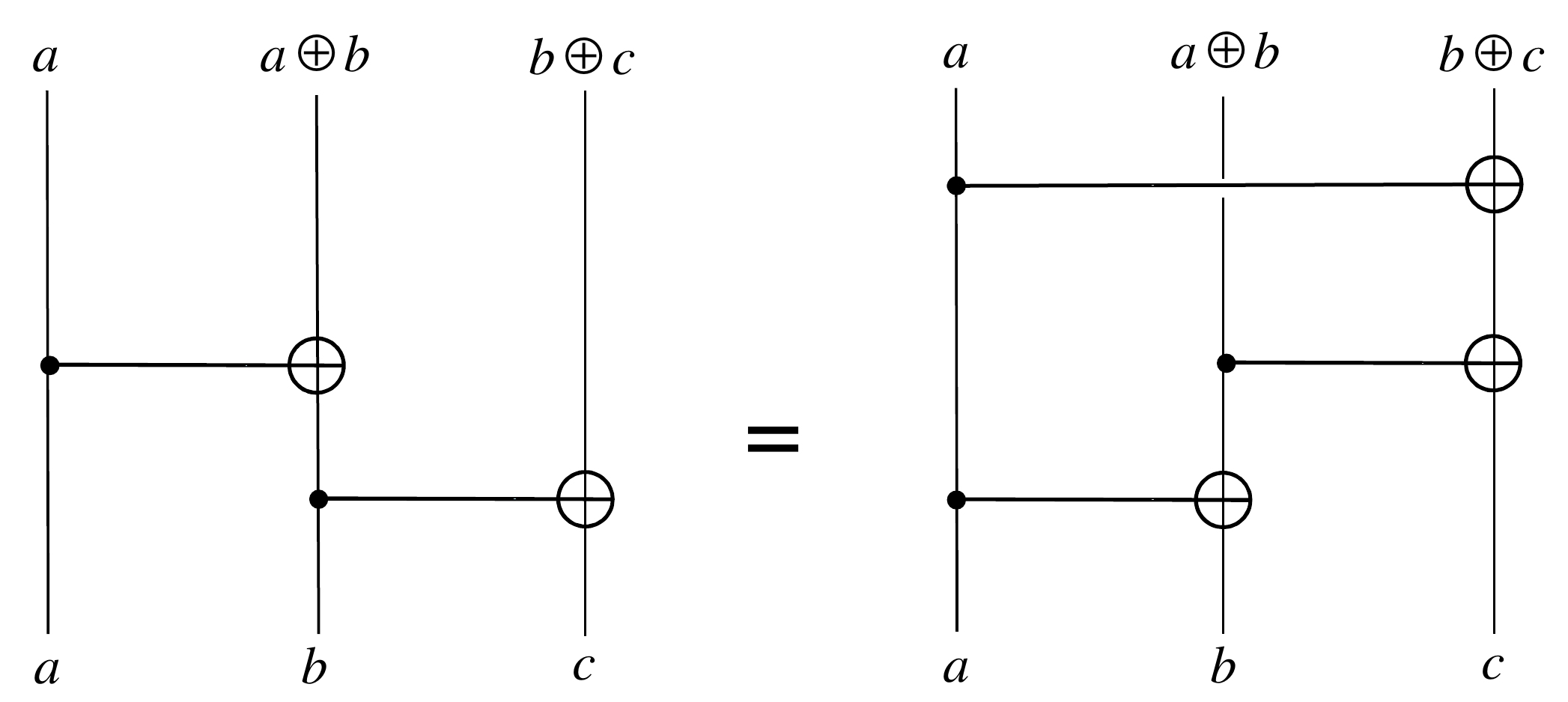}
\end{center}
\caption{The quantum circuit identity used to establish the equivalence of the two circuits shown in Fig.~\ref{fig:brenda-first}.}
\label{fig:cnot-identity}
\end{figure}

As shown in Fig.~\ref{fig:brenda-first}, we can use the quantum circuit identity in Fig.~\ref{fig:cnot-identity} to transform the circuit in which BRenda measures first to an equivalent circuit in which ABby measures first, but this equivalent circuit contains an additional CNOT gate from ABby's meter to BRenda's meter which maximally entangles the two meters. Hence when BRenda measures first, BRenda's and ABby's meters, each considered individually, become maximally mixed; therefore the readout of the meter yields a random outcome. The expected outcome $Z\otimes Z=1$ or $X\otimes X = 1$ is obtained with probability 1/2 rather than probability 1. 

How should we interpret this failure? Has the system measured by each party been modified due to the measurement performed by the other party? Or is it that the measurement meter used by each party, rather than the measured system, has been disturbed by the other party's action? We believe the correct interpretation is that the measurement meters used by BRenda and ABby become entangled when BRenda measures before ABby, while the measured systems themselves are not actually modified. Indeed, we can wait until ABby's (scrambled) meter is emitted in the Hawking radiation, then unite ABby's meter with BRenda's and execute a CNOT gate from ABby's meter to BRenda's before reading out ABby's meter in the $X$ basis and BRenda's meter in the $Z$ basis. In that case both verifications succeed with probability 1. Furthermore, if BRenda measures first, followed by ABby, then BRenda can measure multiple times after ABby's interaction with $B$, each time with a fresh meter. When she does so, although BRenda's first measurement, performed before ABby's, yields a random outcome, all of Brenda's later measurements, performed after ABby's, successfully verify $Z_B\otimes Z_R = 1$ with probability 1. BRenda naturally concludes that ABby's measurement disturbed BRenda's first meter, but had no effect on the state of $BR$ or on the meters used in BRenda's later measurements.

\begin{figure}[t]
\begin{center}
\includegraphics[width=0.5\textwidth]{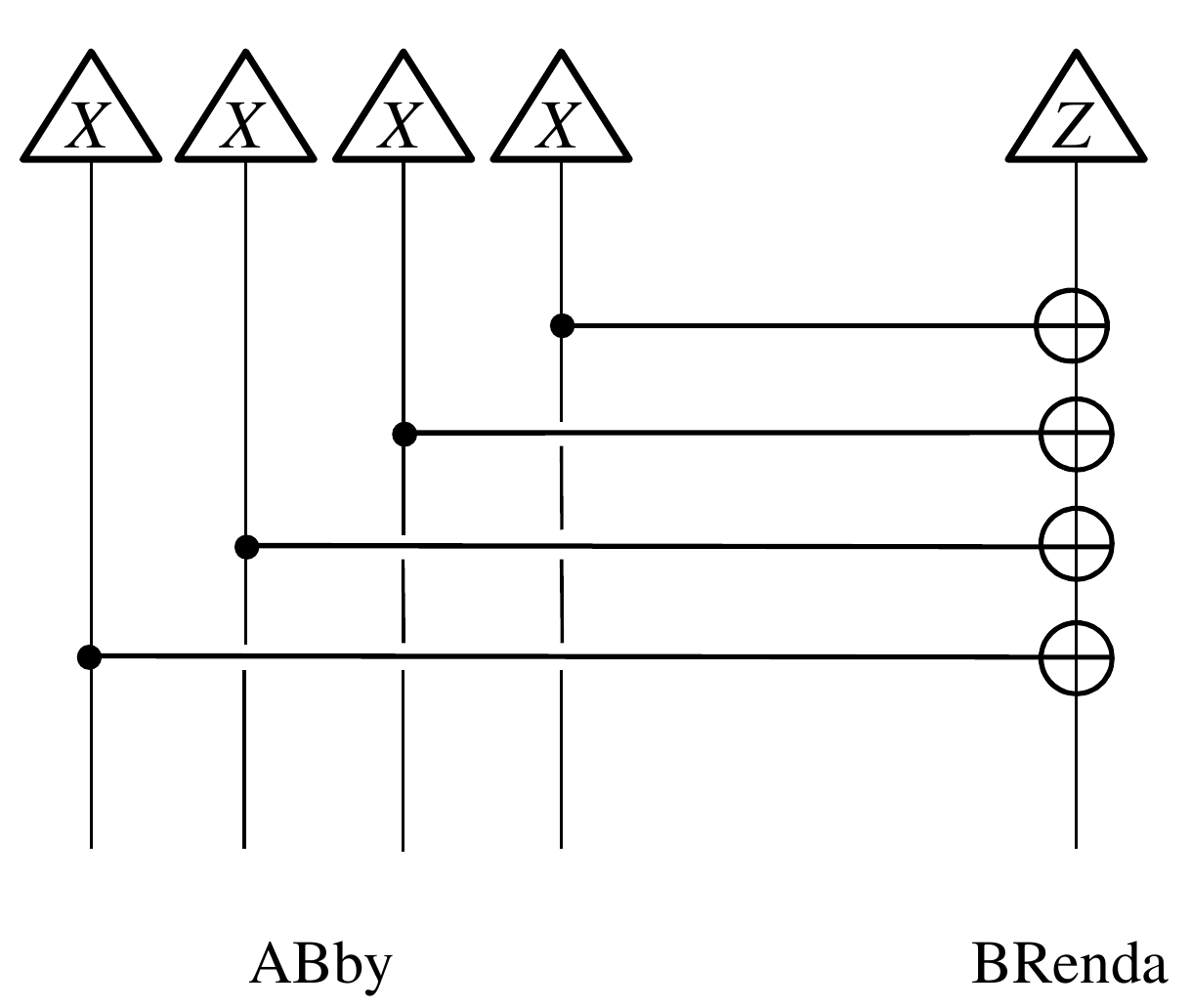}
\end{center}
\caption{If ABby measures $X_A\otimes X_B$ four times after BRenda measures $Z_B\otimes Z_R$ once, the measurements are completed by performing CNOT gates from each of ABby's meter qubits to BRenda's meter qubit, before measuring ABby's qubits in the $X$ basis and BRenda's qubits in the $Z$ basis.}
\label{fig:decode-ghz}
\end{figure}

Likewise, ABby might measure $X_A\otimes X_B$ $k$ times, each time with a new meter, after BRenda measures $Z_B\otimes Z_R$. To see what happens, we may apply the circuit identity Fig.~\ref{fig:cnot-identity} $k$ times, once for each time one of ABby's CNOT gates moves from after BRenda's CNOT to before BRenda's CNOT gate. In the equivalent circuit we obtain, all of ABby's measurements occur before BRenda touches $B$, but the circuit also includes $k$ additional CNOTs acting on the meters, one from each of ABby's meter qubits to BRenda's meter qubit. Thus, after all of ABby's meter qubits are decoded from the Hawking radiation outside the black hole, the quantum state of ABby's $k$ meter qubits and BRenda's meter qubit is
\begin{equation}
\frac{1}{\sqrt{2}}\left( |+\rangle^{\otimes (k+1)} + |-\rangle^{\otimes (k+1)}\right).
\end{equation}
In this state, the marginal density operator of each meter qubit is maximally mixed; if any of the qubits were measured in the $X$ or $Z$ basis the outcome would be uniformly random. To complete the verifying measurements properly, we must perform a compensating CNOT from each of ABby's $k$ meter qubits to BRenda's meter qubit in order to disentangle the meters, and only then measure each of ABby's meters in the $X$ basis and BRenda's meter in the $Z$ basis, as in Fig.~\ref{fig:decode-ghz}. Using this procedure, all the verifying measurements succeed with certainty.

\begin{figure}[t]
\begin{center}
\includegraphics[width=0.4\textwidth]{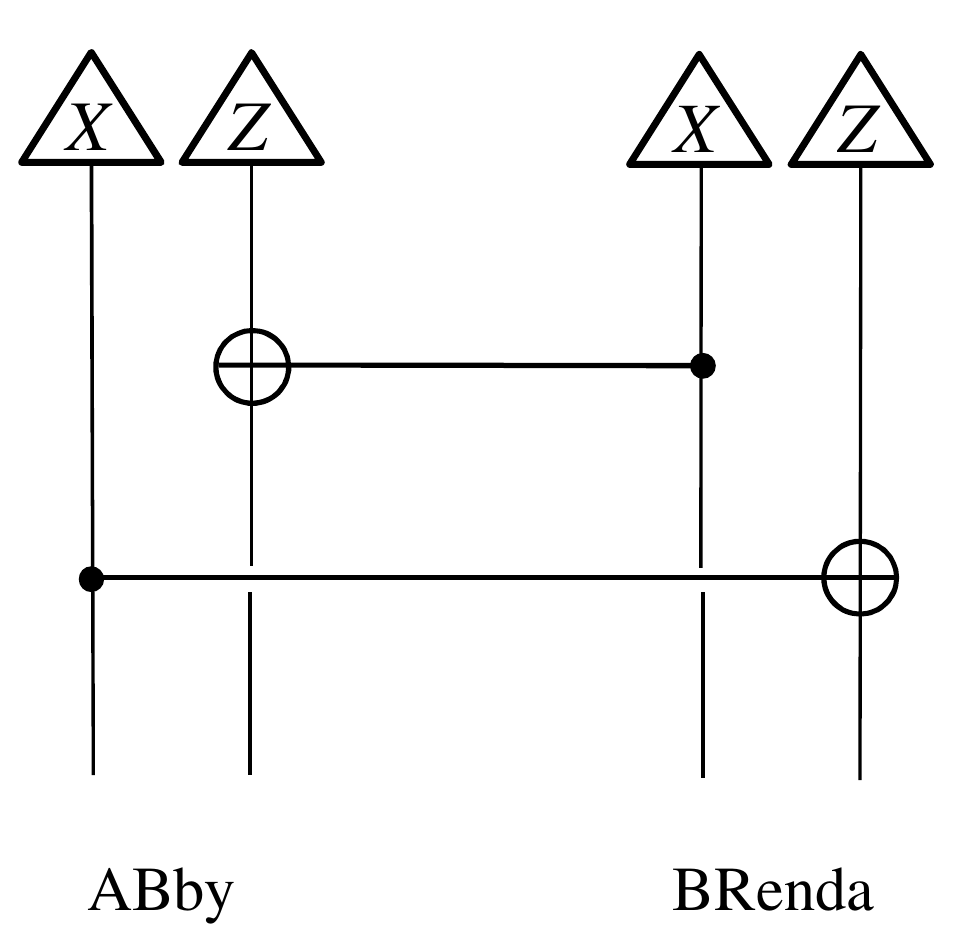}
\end{center}
\caption{If ABby performs a complete Bell measurement after BRenda performs a complete Bell measurement, the measurements are completed by performing this circuit on ABby's and BRenda's meters.}
\label{fig:decode-full-bell}
\end{figure}

This discussion requires only minor modification when both BRenda and ABby perform complete Bell measurements. Now both parties carry two-qubit meters, and using circuit identities we  may again transform the scenario where BRenda measures first to the one where ABby measures first, this time by moving two of ABby's CNOT gates through two of Brenda's. This procedure generates two additional CNOT gates, one from ABby's $X$ meter to BRenda's $Z$ meter and one from Brenda's $X$ meter to ABby's $Z$ meter; these gates maximally entangle ABby's meter and BRenda's. Hence each meter, individually, is maximally mixed, and for both parties the verification of the entangled state $|\phi^+\rangle$ succeeds with only probability 1/4. However, once ABby's meter has been emitted in the Hawking radiation, the meters can be disentangled by reversing these CNOTs, and then measuring in the appropriate basis, as in Fig.~\ref{fig:decode-full-bell}. These disentangled measurements successfully verify the $BR$ and $AB$ entanglement with certainty.

\subsection{Acausal signaling?}

This scenario indicates how postselection might resolve the AMPS puzzle, because verifiable $AB$ entanglement (and hence a smooth black hole horizon) can be reconciled with verifiable $BR$ entanglement (and hence unitarity of the black hole S-matrix) in the HM model. Nevertheless the implications of the scenario are perplexing, because the outcome of a $BR$ measurement performed by BRenda today can depend on whether or not ABby decides to measure $AB$ tomorrow. If BRenda reads her meter immediately after interacting with $B$, the outcome of her Bell measurement yields $|\phi^+\rangle$ if ABby does not measure later on, and is uniformly random if ABby does measure later on. Thus ABby can send a signal into her backward light cone which BRenda may receive. 

Though ABby sends her message by manipulating $A$ behind the horizon, a protocol for backward signaling can be executed outside the horizon. ABby can program a robot to perform Bell measurement on $AB$, and then signal BRenda by either holding the robot or dropping it into the black hole. Of course, if black hole evaporation is unitary, then quantum information manages to escape from behind the black hole horizon, which is already in a sense ``causality violating,'' but the postselection in the HM model potentially allows a stronger type of causality violation, detectable by observers who stay outside the black hole.

There is a limit on how far backward in time the signal can propagate, if we are to describe this protocol semiclassically. After BRenda interacts with $B$, ABby's meter should enter the black hole within less than the scrambling time $O(m \log m)$ (where $m$ is the black hole mass), if it is to ``catch up'' with $B$'s Unruh partner $A$ in a region of low curvature inside the black hole. On the other hand, we may chain together many such protocols; if ABby can send a signal which is received by BRenda a time $t$ before ABby sends it, then $N$ parties acting together should be able to send a signal which is received time $Nt$ before it is sent (ignoring the time needed for each party to execute her part of the protocol, which in principle can be parametrically small compared to the scrambling time). Then the only fundamental limitation on how far backward a signal can be sent is the black hole lifetime $O(m^3)$, which might be further extended by feeding the black hole a steady diet of infalling matter to maintain its mass as it radiates. 

One way to prevent acausal signaling would be to place further restrictions on ABby's actions behind the horizon, going beyond the enforcement of unitarity. In particular, if ABby is unable to apply the unitary transformation to the Unruh partner of $B$, or equivalently if this transformation is reversed by the final-state boundary condition, then ABby will not be able to alter the outcome of a measurement BRenda performed previously. On the other hand, ABby will also be unable to verify the $AB$ entanglement --- she will perform a measurement on $B$ alone, rather than a joint measurement on $AB$ (a measurement which, by the way, will damage the entangled state of $BR$). 

This way of enforcing causality in the HM model severely limits ABby's ability to measure the black hole interior, or at any rate limits her ability to create a measurement record that can survive after the black hole evaporates. If no such measurements were possible, it might be reasonable to claim that the black hole interior does not really exist, or in other words that the horizon is really a singular firewall.

Another possible way to prevent acausal signaling is to invoke the complexity of decoding the Hawking radiation \cite{harlow-hayden}. To decipher ABby's message, BRenda needs to perform a joint measurement on $BR$; before that she needs to extract $R$. In principle, she could decode $R$ well before she touches $B$ to initiate the protocol; the decoding time, then, could be comparable to the $O(m^3)$ Page time, and long compared to the scrambling time. Even so, high decoding complexity may make the signaling protocol infeasible. 

We expect, at least in the case of an asymptotically AdS bulk spacetime, to be able to describe the dynamics of an evaporating black hole using a dual field theory defined on the boundary of spacetime, where the field theory is unitary and local \cite{ads-cft}.  Acausal signaling outside the black hole horizon, if it can occur, might be hard to reconcile with strict causality in the field theory. (Local observables used in the bulk signaling protocol would correspond to precursor operators, which might be highly nonlocal, in the boundary theory \cite{precursor}.) On the other hand, since the signaling protocol involves decoding the Hawking radiation, the corresponding process in the dual field theory might be astoundingly complex. One possibly consistent point of view is that by the time the decoding is completed either the black hole has completely evaporated or a firewall has arisen, preventing signaling from inside the black hole. 

Bousso and Stanford \cite{bousso-stanford} have made observations related to ours regarding measurements inside black holes, albeit from a different perspective and in a different language. 

\subsection{Computational power of the HM model}

If acausal signaling in the bulk spacetime really is possible, it is natural to wonder whether ``time-travel paradoxes'' threaten the consistency of the theory. We emphasize, though, that if we demand overall unitarity of the black hole S-matrix (or of the boundary field theory), the allowed acausal bulk phenomena are highly constrained and not obviously inconsistent. 

Another natural question concerns the computational complexity of simulating the HM model with a quantum computer. Quantum computation with final-state projection is known to be PP-complete  \cite{aaronson}. Hence general final-state projection models allow very hard computational problems to be solved ``efficiently'' (in particular, PP contains the complexity class NP, the class of problems for which a solution can be efficiently verified using a classical computer). 

It is therefore important to recognize that the HM model admits only a restricted kind of postselection, if we require the model to be compatible with unitarity. Though it is still an open question whether quantum gravity can be simulated efficiently with a standard quantum computer, so it is at least possible in principle based on current knowledge that quantum gravity computers can solve problems which are beyond the reach of standard quantum computers, we know no reason why the computational power of the (unitary) HM model should exceed that of other quantum gravity models.

\section{Discussion}
The AMPS puzzle has deepened the mystery surrounding the fate of quantum information that falls into a black hole. AMPS investigated the compatibility of three reasonable assumptions: (1) unitarity of black hole evaporation, (2) smoothness of the black hole event horizon, and (3) validity of local effective field theory outside a black hole. They argued that these three assumptions are inconsistent, since together they imply that quantum correlations can be polygamous, contrary to standard quantum mechanics. 

Our main point is that quantum correlations can be polygamous in the Horowitz-Maldacena final-state projection model, permitting these three assumptions to be reconciled. In the HM model, quantum information escapes from the black hole interior via postselected quantum teleportation, due to a boundary condition imposed at the spacelike singularity. Loosely speaking, quantum information flows forward in time from past infinity to the singularity, backward in time from the singularity to the horizon, then forward in time from the horizon to future infinity. If suitable dynamical constraints are satisfied, this flow of information is essentially equivalent to a manifestly unitary causally ordered flow moving only forward in time, at least for the purpose of describing the viewpoint of observers who stay outside the black hole. These constraints are nearly fulfilled by generic dynamical models, but as best we can tell they can be rigorously fulfilled only by fine tuning the model. On the other hand, since the HM model is formulated on a semiclassical spacetime background, achieving unitarity up to exponentially small corrections using a generic final-state boundary condition might be regarded as a success of the model.

In the HM model, observables inside the horizon fail to commute with observables outside the horizon acting on the same time slice, because in the corresponding causally ordered information flow, the outside observables act on the same system as the inside observables, but at a later ``time.'' Other features of black hole complementarity are also realized; in particular, from the viewpoint of an observer who stays outside, the black hole behaves like a rapidly scrambling quantum system interacting with its surroundings. 

If a black hole has an interior, it should be possible to perform measurements inside the black hole. Furthermore, records of such measurements should in principle be recoverable from the Hawking radiation emitted after the black hole evaporates. We have analyzed such measurements in the HM model, finding that actions performed behind the black hole horizon can enable causality-violating signaling outside the horizon. 

However, this signaling protocol only works if Hawking radiation can be rapidly decoded; the high computational complexity of decoding \cite{harlow-hayden} may inoculate the HM model against acausal signaling outside the horizon. In that event, we know no reason why the physics outside the horizon could not be accurately captured by a dual boundary field theory as in AdS/CFT duality, where the field theory is unitary and local.  The truly novel physics of the HM model occurs inside the black hole, particularly at the singularity; the model may provide helpful hints about how a dual description of the black hole interior should work, if such a description exists.

In postselected quantum mechanics, cloning of quantum states is possible, and because monogamy of quantum entanglement can be relaxed, we know no logically compelling argument for the existence of a firewall at the black hole horizon within the context of the HM model; conceivably, though, the horizon could nevertheless fail to be smooth for reasons other than those originally promulgated by AMPS. (See \cite{marolf,bousso1,bousso2} for other arguments supporting the existence of firewalls.)

Like all other resolutions of the AMPS puzzle proposed so far, the HM model will need to be developed further before it can be conclusively assessed. In particular, we should strive to expunge the dynamical fine tuning the model seems to require, or to explain persuasively why the fine tuning is somehow natural. 

Even if the HM model turns out to be wrong in detail, we believe that the picture of information flow in black hole spacetimes provided by the model is interesting and valuable. This picture reminds us that the global physics of the black hole interior could be subtle, and in particular that fundamental properties of standard quantum mechanics such as the no-cloning principle and monogamy of entanglement might be relaxed in a complete theory of quantum gravity. And if nature really indulges in postselection at future spacelike singularities, we may anticipate deep consequences in quantum cosmology as well as black hole physics.

\vskip .5cm
\begin{center} {\bf Acknowledgments} \end{center}
We gratefully acknowledge very valuable discussions with Alexei Kitaev. SL thanks Max Tegmark for helpful discussions, and JP thanks Raphael Bousso, Daniel Harlow, Juan Maldacena, Joe Polchinski, and Douglas Stanford for inspiring discussions and correspondence regarding final-state projection models. JP appreciates many helpful interactions with other participants at the August 2013 KITP workshop ``Black Holes: Complementarity, Fuzz, or Fire,'' and we also benefited from comments on the manuscript from Don Marolf and Douglas Stanford. The research of SL was supported in part by DARPA, by AFOSR, by the ARO under a MURI program, and by Jeffrey Epstein. The research of JP was supported in part by NSF, ARO, and DOE. The Institute for Quantum Information and Matter (IQIM) is an NSF Physics Frontiers Center with support from the Gordon and Betty Moore Foundation.

\end{document}